\documentclass[8.5pt,twoside,twocolumn]{article}
\usepackage[super,sort&compress,comma]{natbib} 

\usepackage[version=3]{mhchem} 
\usepackage[T1]{fontenc}

\usepackage{booktabs}
\usepackage{color}
\usepackage{graphicx}
\usepackage{lscape}
\usepackage{rotating}
\usepackage{dcolumn}
\usepackage{bm}
\usepackage{dcolumn}
\usepackage{etoolbox}
\usepackage{booktabs}
\usepackage{multirow}
\usepackage[table,xcdraw]{xcolor}
\usepackage{tabularx}
\usepackage{lipsum}
\usepackage{amsmath}
\usepackage{braket}

\usepackage{footnote}

\makesavenoteenv{table}

\newcommand{\RNum}[1]{\uppercase\expandafter{\romannumeral #1\relax}}

\usepackage{sidecap}

 \usepackage{adjustbox}
 \usepackage{breqn}
 \usepackage{array,multirow}
 \usepackage{comment}

\usepackage{mhchem}
\usepackage{times,mathptmx}
\usepackage{sectsty}
\usepackage{balance}

\usepackage{amsmath,graphicx} 
\usepackage{lastpage}
\usepackage[format=plain,justification=raggedright,singlelinecheck=false,font=small,labelfont=bf,labelsep=space]{caption} 
\usepackage{fancyhdr}
\usepackage{graphicx}
\usepackage{etoolbox}

\usepackage{mypackages}  
\usepackage[justification=justified]{caption}
\usepackage{verbatim} 
\usepackage{lmodern}

\usepackage{cleveref}

\begin{document}


\thispagestyle{plain}

\renewcommand{\thefootnote}{\fnsymbol{footnote}}
\renewcommand\footnoterule{\vspace*{1pt}%
\hrule width 3.4in height 0.4pt \vspace*{5pt}} 
\setcounter{secnumdepth}{5}

\makeatletter 
\def\subsubsection{\@startsection{subsubsection}{3}{10pt}{-1.25ex plus -1ex minus -.1ex}{0ex plus 0ex}{\normalsize\bf}} 
\def\paragraph{\@startsection{paragraph}{4}{10pt}{-1.25ex plus -1ex minus -.1ex}{0ex plus 0ex}{\normalsize\textit}} 
\renewcommand\@biblabel[1]{#1}            
\renewcommand\@makefntext[1]%
{\noindent\makebox[0pt][r]{\@thefnmark\,}#1}
\makeatother 
\renewcommand{\figurename}{\small{Fig.}~}
\sectionfont{\large}
\subsectionfont{\normalsize} 

\renewcommand{\headrulewidth}{1pt} 
\setlength{\arrayrulewidth}{1pt}
\setlength{\columnsep}{6.5mm}
\setlength\bibsep{1pt}

\twocolumn[
  \begin{@twocolumnfalse}
    \noindent\LARGE{\textbf{Large yet bounded: Spin gap ranges in carbenes}}
    
\vspace{0.6cm}

\noindent\large{\textbf{Max Schwilk,\textit{$^{a,b}$} Diana N. Tahchieva,\textit{$^{a,b}$}  O. Anatole von Lilienfeld,\footnotemark\textit{$^{a}$} }}\vspace{0.5cm}



\vspace{0.6cm}

\noindent \normalsize{
Despite its relevance for chemistry, the electronic structure of free carbenes throughout 
chemical space has not yet been studied in a systematic manner. 
We explore a large and systematic carbene chemical space 
consisting of eight thousand diverse and common carbene scaffolds in their singlet and triplet state computed at controlled accuracy (higher order multireference level of theory) and with verified carbene character in the electronic structure. 
Originating in strong electron correlation, a hard upper 
limit for the singlet-triplet gap is found to emerge 
at around 30 kcal/mol
for all the carbene classes in this chemical space.
We also observe large vertical and adiabatic spin gap ranges within many carbene classes ($>$100 and $>$60 kcal/mol, respectively), and we report novel relationships between compositional, structural, and electronic degrees of freedom. 
Our QMspin data base includes numerical results for $\approx$13'000 MRCI calculations on randomly selected carbene scaffolds.


}
\vspace{0.5cm}
 \end{@twocolumnfalse}
  ]



\footnotetext{\textit{$^{a}$~Institute of Physical Chemistry and National Center for Computational Design and Discovery of Novel Materials (MARVEL), Department of Chemistry, University of Basel, Klingelbergstrasse 80, CH-4056 Basel, Switzerland, E-Mail: anatole.vonlilienfeld@unibas.ch. \\
$^{b}$~M.\ S.\ and D.\ N.\ T.\ contributed equally to this work.}}

For more than half a century carbenes have been known for their key role as transient intermediates in a variety of organic chemistry reactions.\cite{Woodworth1959JACS,csizmadia1968mechanism}
Due to their high reactivity, carbenes often have a short lifetime and are therefore most commonly formed \emph{in situ}.\cite{Chu2016JACS} 
Carbenes with a lifetime longer than typical reaction intermediates offer the perspective of more versatile applications.
Therefore, an important step in carbene chemistry was the synthesis of stable and bottleable singlet state carbenes.\cite{arduengo1991stable,arduengo19981,igau1988analogous} 
Triplet carbenes, on the other hand, may have lifetimes of at most few hours 
and even such ``persistent'' triplet carbenes are highly reactive. 
Highly reactive carbenes may be experimentally accessible only via isolation in a rare gas matrix at a few  Kelvin\cite{sander1993carbenes} 
and carbene properties at such experimental conditions can be directly compared to results from quantum chemistry calculations.
As such, the high reactivity of many carbenes makes experimental observations \cite{modarelli1991interception} 
challenging and carbene characterization\cite{droege2010angew} has led to particular fruitful cooperations with computational approaches.\cite{shavitt1985geometry,gerbig2013computational} 
Despite their experimental importance, only few computational studies have covered diverse carbene classes.\cite{gronert2011stabilities, mieusset2008carbene, vasiliu2017characterization,shainyan2013carbenes} 
More comprehensive carbene studies involving subspaces of chemical space with more than a hundred compounds have not yet been published, indicating that our understanding of trends in carbene chemistry are far from complete.
Ultimately, electronic structure validation at controlled accuracy of combinatorically derived carbene structures can enable the exploration of new relationships which could prove helpful for their use in materials discovery and design efforts.\\
\indent Here, we report on discoveries made based on an extensive analysis of a large set of newly computed accurate structures and spin splittings at higher order multireference level (MR) of theory [MR configuration interaction (MRCI)] for eight thousand carbenes which we generated from a random subset of molecular scaffolds from supposedly stable and synthetizable organic molecules with up to nine heavy atoms of the elements H, C, N, O, and F (QM9 data set).\cite{ramakrishnan2014quantum}
Surprisingly, for almost 90\% of these initial carbene candidates we could confirm that the divalent carbon center always has two well-localized non-bonding orbitals, thereby qualifying as  ``genuine'' carbenes.
The obtained data set is named QMspin and contains numerous carbene classes as they emerge from the chemical space of QM9. 
Our analysis elucidates the combinatorial possibilities and limitations of electronic carbene spin state design.
\\
\\
%
%
     \begin{figure*}[t!]
        \centering
         \includegraphics[width=0.95\linewidth]{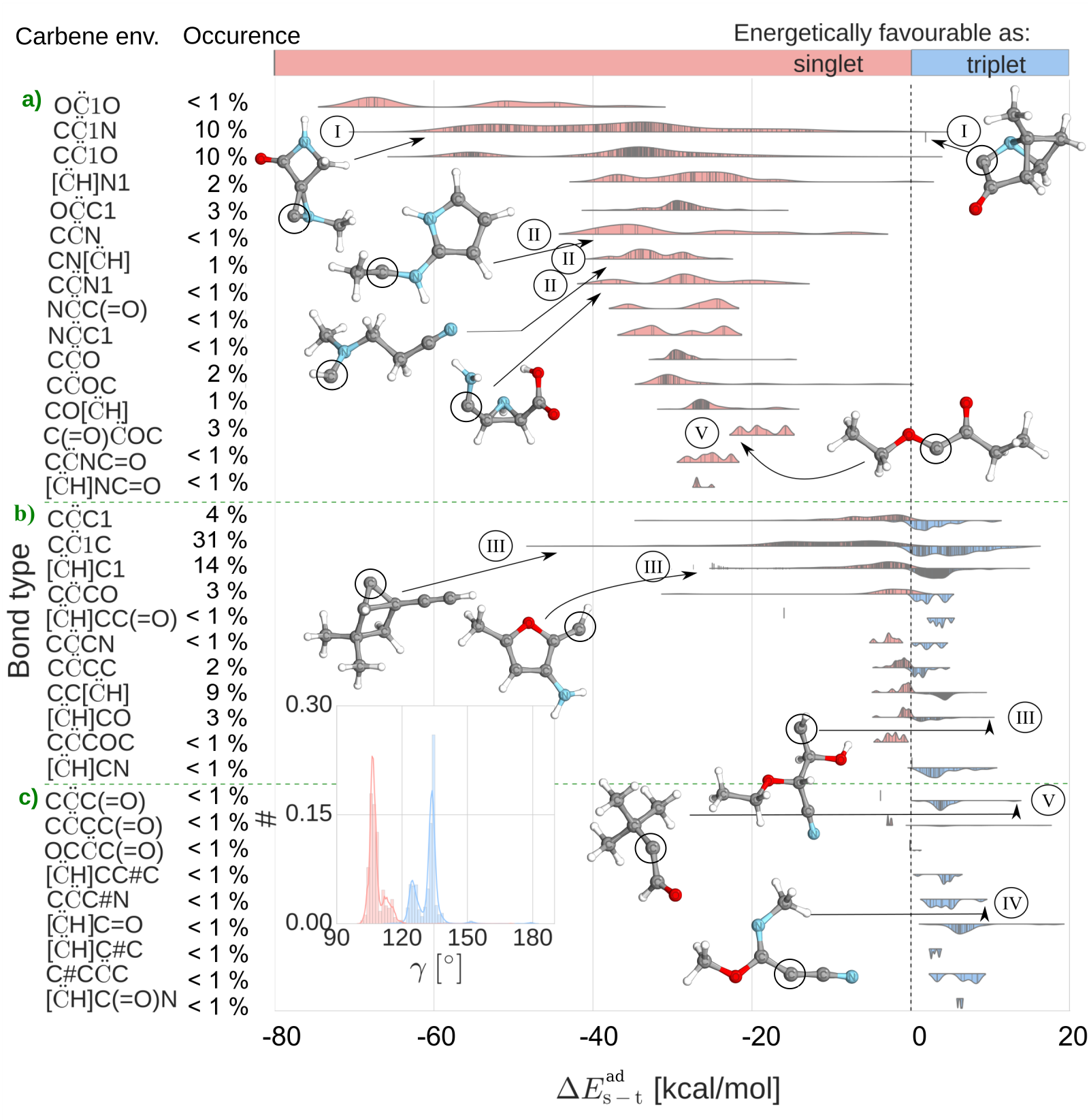}
        \caption{Distribution of the 5021 adiabatic singlet-triplet gaps computed with MRCI for all functional groups with more than 0.1\% occurrence in the QMspin data set. Carbene classes are grouped according to their main characteristics (a)-(c) (see Methods section for details). $\Delta E_{\rm{s-t}}^{\rm{ad}} > $ 0 (blue) and $\Delta E_{\rm{s-t}}^{\rm{ad}} <$ 0 (red) correspond to carbenes with a singlet and triplet ground state, respectively. R'-$\ddot{\rm{C}}$1-R and R'-$\ddot{\rm{C}}$-R1 denote carbenes in ring and ring in $\alpha$ position, respectively. A selection of extremal carbenes in terms of $\Delta E_{\rm{s-t}}^{\rm{ad}}$ are shown as insets with their triplet state structure. Roman numerals indicate examples of the effect of the carbene environment described in the text. As an inset plot is given the distribution of carbene angle for singlet (red) and triplet (blue) state geometries (only acyclic carbenes).}
        \label{fig:st_stats}
    \end{figure*}
%
%
\textbf{Results and Discussion}
\\
\\
\indent \textbf{The versatility of adiabatic spin gaps.}
A coarse classification of the carbenes in our QMspin data set can be made into carbenes with (a)~$\sigma$-electron-withdrawing and $\pi$-electron-donating $\alpha$-substituents, e.\ g.\ oxo- and amino-carbenes (b)~aliphatic and aromatic $\alpha$-substituents, e.\ g.\ vinyl-,\cite{barluenga20088+} alkynyl-,\cite{seburg2009structure} alkyl-,\cite{lavallo2004synthesis} and aryl-\cite{cattoen2004amino}carbenes, and (c)~$\alpha$-substituents with moderate $\sigma$-electron- and $\pi$-electron-withdrawing character, such as cyano-\cite{freeman2006singlet} and carbonyl-carbenes.
In general, group (a) strongly stabilizes the singlet state\cite{bourissou2000stable,tomioka1997persistent} while group (b) stabilizes the triplet state or has a near zero spin gap.\cite{hirai2009persistent}
For carbonyl-carbenes, which are part of group (c), strong conformational changes can arise between the singlet and triplet state and especially high spin gaps may result.\cite{toscano1994carboethoxycarbene,scott2001singlet,kim1980geometric}\\  
\indent Figure \ref{fig:st_stats} shows the distribution of the adiabatic spin gap $\Delta E_{\rm{s-t}}^{\rm{ad}}$ for the different combinations of $\alpha$-substituents, given in SMILES\cite{w1988}
-like abbreviations, referred to in typewriter font in the following. 
The carbenes are grouped into the categories (a)--(c) for which the expected main trends are clearly reproduced: Group (a) carbenes mostly have spin gaps between -50 and -20 kcal/mol, group (b) carbenes mostly distribute within 10 kcal/mol around a zero spin gap, and group (c) carbenes can reach one of the highest adiabatic spin gaps that are observed (around 10 kcal/mol).
Hence, the here represented chemical space, whose molecular scaffolds were constructed by purely combinatorial rules, elucidates the same main trends as observed by experimental approaches or theoretical studies on manually selected representative carbenes.
However, also extreme molecules in terms of spin gaps emerge from the data set and some of them are shown as insets in the figure.
%
%
We observe broadest spreads of $\Delta E_{\rm{s-t}}^{\rm{ad}}$ for cyclic carbenes [\texttt{R-$\ddot{\rm{C}}$1-R} in Figure \ref{fig:st_stats}] mainly due to the geometrical constraints imposed by the ring structure.
The most remarkable trends and relationships can be grouped as follows (inset molecules in Fig.\ \ref{fig:st_stats} are labeled with the roman letter accordingly):

I. Two oxy-substituents, occuring for cyclic carbenes [\texttt{(O-$\ddot{\rm{C}}$1-O}) in the Figure], show the lowest average spin gap ($\approx$-60 kcal/mol).
The spin gaps of cyclic carbenes with one oxy-substituent [\texttt{O-$\ddot{\rm{C}}$1-O} in the Figure] distribute around an average value that is $\approx$20 kcal/mol higher, providing statistical evidence for an ``additivity'' of this $\sigma$-withdrawing functional group in stabilizing the singlet state. 
Acyclic hydroxy- (\texttt{C$\ddot{\texttt{C}}$O}, \texttt{O$\ddot{\texttt{C}}$C1}) and oxy-carbenes (\texttt{C$\ddot{\texttt{C}}$OC}, \texttt{C$\ddot{\texttt{C}}$OC(=O)}) show an average spin gap of $\approx$-30 kcal/mol.
Carbenes with N-substituents in $\alpha$, to a large part comprised of cyclic alkyl aminocarbenes (cAAC)\cite{soleilhavoup2014cyclic}  [\texttt{C-$\ddot{\rm{C}}$1-N} in Fig.\ \ref{fig:st_stats}], span a previously unreported wide range of adiabatic spin gaps.  
For example, a spiro-aziridine-derived carbene has $\Delta E_{\rm{s-t}}^{\rm{ad}}$=-60.6 kcal/mol 
while we also find a cAAC with a positive spin gap (see inset in the Figure).
The latter is the first reported cAAC with an expected triplet ground state: The bicyclic structure imposes a quasi-planar conformation on the O=C-$\ddot{\rm{C}}$-N axis and the resulting strong $\pi$-interaction favours the triplet state. 


II. Acyclic amino-carbenes constitute the groups \texttt{CN$[\ddot{\texttt{C}}\texttt{H}]$}, \texttt{C$\ddot{\texttt{C}}$N}, \texttt{N$\ddot{\texttt{C}}$C1}, and part of the compounds in \texttt{C$\ddot{\texttt{C}}$N1}, \texttt{[$\ddot{\texttt{C}}$H]N1}. 
They show mostly a singlet ground state in accordance with other computational results in the literature,\cite{gronert2011stabilities} the lowest values reach $\Delta E_{\rm{s-t}}^{\rm{ad}}\approx$-40 kcal/mol (see inset example). 
Acyclic alkyl amino carbenes (acAACs) have gained considerable interest in catalysis in recent years,\cite{martin2011stable} here a wide range of adiabatic spin gaps for these compounds is reported for the first time. 

III. For alkyl- and aryl-carbenes exceptionally low spin gaps ($\Delta E_{\rm{s-t}}^{\rm{ad}}<$-20 kcal/mol) arise for $\pi$-donor substituents, such as furan derivatives, or in the presence of strong ring strains, as shown in inset examples. 
These carbenes comprise the following groups, distinguished by their functional groups in $\beta$-position: \texttt{[$\ddot{\texttt{C}}$H]CC\#C}, \texttt{CC[$\ddot{\texttt{C}}$H]}, \texttt{[$\ddot{\texttt{C}}$H]CC(=O)}, \texttt{C$\ddot{\texttt{C}}$CO}, \texttt{C$\ddot{\texttt{C}}$CC}, \texttt{[$\ddot{\texttt{C}}$H]CO}, \texttt{C$\ddot{\texttt{C}}$COC}, \texttt{[$\ddot{\texttt{C}}$H]CN}, \texttt{C$\ddot{\texttt{C}}$CC(=O)} and some of the systems in \texttt{[$\ddot{\texttt{C}}$H]C1} and \texttt{C$\ddot{\texttt{C}}$C1}.  

IV. In accordance with previous results\cite{gronert2011stabilities} the cyano (\texttt{C$\ddot{\texttt{C}}$C\#N}) and alkynyl (\texttt{C\#C$\ddot{\texttt{C}}$C, [$\ddot{\texttt{C}}$H]C\#C}) functional groups stabilize the triplet state (one inset example is given in the Figure). 

V. One of the highest values for $\Delta E_{\rm{s-t}}^{\rm{ad}}$ ($\approx$ 14 kcal/mol) are obtained for carbonyl-carbenes (\texttt{C$\ddot{\texttt{C}}$C(=O)} and \texttt{[$\ddot{\texttt{C}}$H]C=O})
when the carbonyl group is part of an electrophilic $\pi$-system or steric constraints enforce a large bond angle at the carbene center (see inset example). 
However, inspection of the spin gap distributions of ``mixed'' group (a) and (c) carbenes [see e.\ g.\ \texttt{C\#C$\ddot{\texttt{C}}$O}, \texttt{C(=O)$\ddot{\texttt{C}}$OC}, \texttt{N$\ddot{\texttt{C}}$C(=O)}, and inset example] puts into evidence that the oxy- or amino-group generally prevails in its influence on the spin gap.
We grouped these carbenes therefore in the class (a). \\ \indent
%
\textbf{Carbene bond angle.}
The relationship between the bond angle at the carbene center ($\angle$ R-$\ddot{\rm{C}}$-R') and the spin gap has always been of strong scientific interest in carbene chemistry.\cite{bourissou2000stable,tomioka1997persistent} 
The basic reasoning is: the smaller the carbene angle, the higher the $s$-character of the energetically low-lying non-bonding orbital, hence the higher the orbital energy splitting and the lower the singlet state energy.\cite{tonner2007bonding} 
The inset plot of Fig.\ \ref{fig:st_stats} shows a surprisingly clear-cut relationship: Almost all acyclic carbene triplet state structures have a carbene center bond angle above 120$^{\circ}$ while for the corresponding singlet state structures this value is almost always below 120$^{\circ}$.
The angle distribution for the singlet state has a peak at around 105$^{\circ}$, while the one for the triplet state distributes around 130$^{\circ}$. \\ \indent 
\textbf{Strong electron correlation.}
Multi-reference methods allow for explicitly measuring the degree of resonance stabilization in the singlet state of the two electrons in the two carbene non-bonding orbitals [(2e,2o)], in our case CASSCF averaged over the two spin states.
Natural orbitals in the (2e,2o) active space describe the singlet ground state by two closed shell configurations with their coefficients $c_R$ and $c_S$ of the non-bonding carbene orbitals $r$ and $s$, whose degree of ``entanglement'' can be given by an angle $\theta=\arctan\big(\frac{|c_S|}{|c_R|}\big)$ with $c_R \ge c_S$.
The values of $\theta$ can then range from 0$^{\circ}$ to 45$^{\circ}$, where an angle of a few degrees indicates weak resonance stabilization and $\theta = 45^{\circ}$ indicates maximum resonance stabilization with two equivalently important configurations in the molecular wave function (see Figure SI-1 
for an illustration of the carbene molecular orbital diagram and the entanglement angle $\theta$).
We find that natural orbitals as eigenfunctions of the one-particle state-averaged density matrix reproduce well the chemical concepts of the carbene non-bonding orbitals' hybridization (indeed mostly $sp^2$- and $p$-orbitals)
and are therefore also a useful qualitative tool for visual inspection.
It should be noted that the orbital entanglement can also be expressed as diradical character of the singlet state by an appropriate pairwise rotation of the active space orbitals and similar measures have been used to quantify the diradical character of the singlet ground state in extended $\pi$ systems.\cite{minami2013signature}
The distribution of the entanglement angles for the singlet and triplet state optimized structures (depicted in Fig.\ SI-2) 
shows that this strong electron correlation generally plays a significant role in singlet state stabilization ($10^{\circ} \le \theta \le 20^{\circ}$ in most cases).
Interestingly, the large values of $\theta$ do not directly correlate with large values of the carbene bond angle.
The latent presence of the strong orbital entanglement in the singlet state has been discussed early on in computational carbene chemistry (see the SI for a discussion of additional details).\cite{shavitt1985geometry}
\begin{figure}
    \centering
    \includegraphics[width=1.04\linewidth]{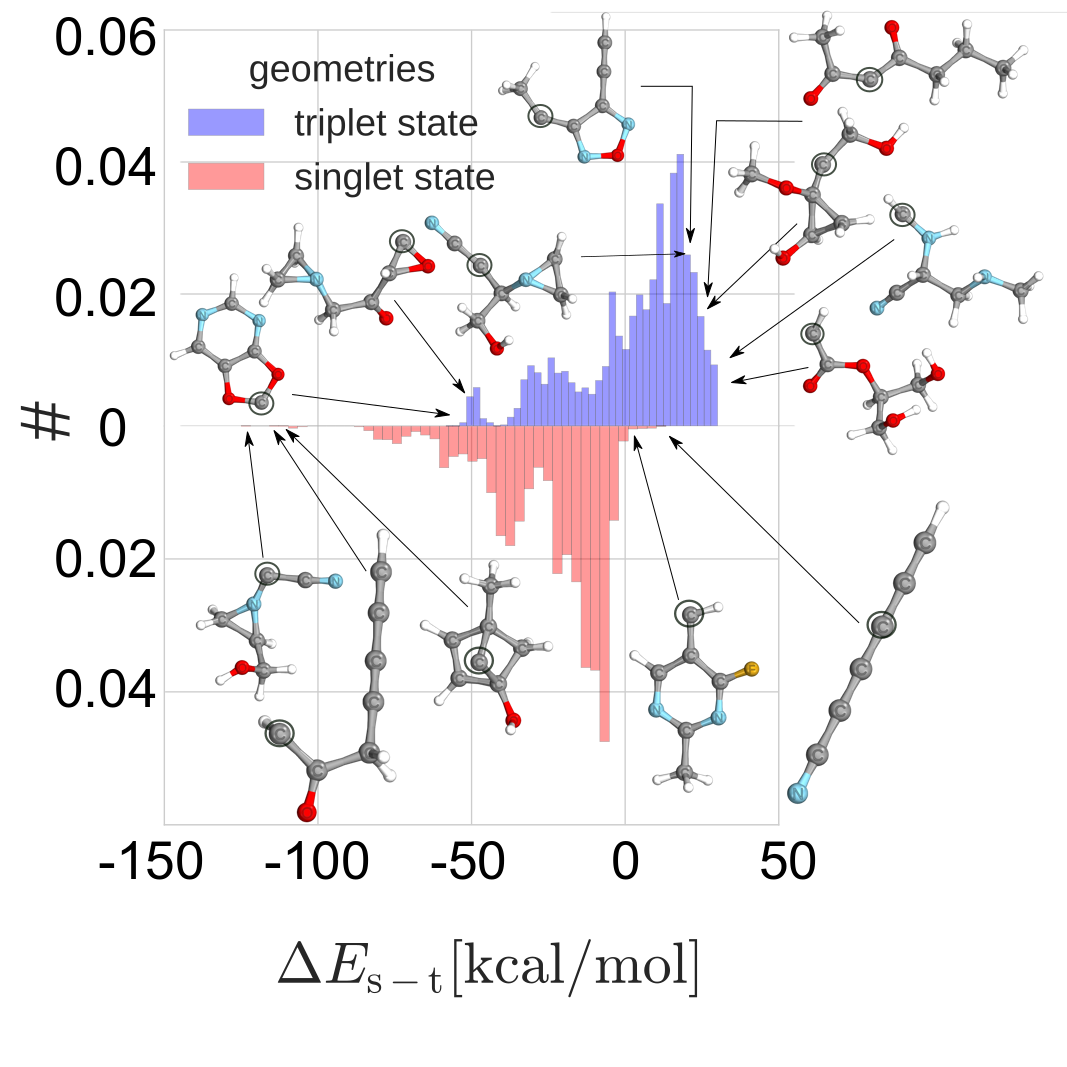}
    \caption{Normalized distribution of $\Delta E_{\rm{s-t}/\!/t}$ (blue) and $\Delta E_{\rm{s-t}/\!/s}$ (red) calculated with MRCISD+Q-F12/cc-pVDZ-F12. Examples for extreme carbenes with lowest singlet and highest triplet are given as inset figures.}
    \label{fig:en_distr}
\end{figure}
\\
\indent \textbf{The versatility of vertical spin gaps.}
The study of vertical spin gaps $\Delta E_{\rm{s-t}}$ plays a key role in analyzing carbene reactivity in the context of spin-flip processes.
Furthermore, vertical spin gaps also have a conceptual value in probing the limits in spin state design imposed by the carbene electronic structure independently from structural relaxation.
In the following, vertical spin gaps of the singlet and triplet optimized structures are named $\Delta E_{\rm{s-t}/\!/s}$ and $\Delta E_{\rm{s-t}/\!/t}$, respectively.
Fig.\ \ref{fig:en_distr} shows the vertical spin gap distributions along with exemplary extreme vertical spin gap cases as inset structures. 
Further extreme vertical spin gap cases are illustrated in Tables\ SI-1 and SI-2. 
For triplet state geometries, the lowest spin gaps occur at specific oxirane- and dioxane-derived cyclic carbenes in the lines of our observations for adiabatic spin gaps.
Unexpected and previously unreported structure--property relationships on the vertical carbene spin gaps are:
\vspace{-0.7em}
\begin{itemize}
\setlength\itemsep{-0.3em}
\item The molecules with the highest vertical spin gaps of the triplet state structure $\Delta E_{\rm{s-t}/\!/t}$ are all in a surprisingly narrow range of 28 to 30 kcal/mol (see Fig.\ \ref{fig:en_distr} and Table\ SI-1). 
These molecules are monosubstituted methylidenes, namely amino-, amido-, cyano-, or carbonyl-carbenes. 
The bond angle at the carbene center in these cases is around 130$^\circ$ and the singlet state shows a strong multi-reference character. 
Disubstitued methylidenes (e.\ g. cyano-, oxo-, methoxy-carbenes) show a similar spin gap range of $20 \le \Delta E_{\rm{s-t}/\!/t} \le 30$ kcal/mol and also for acAACs we observe an unexpectedly large stabilization of the triplet state structure with vertical spin gaps $\Delta E_{\rm{s-t}/\!/t} >$20 kcal/mol  (see Tab.\ SI-1).
\item Fig.\ \ref{fig:en_distr} and Table\ SI-2 
also show a selection of the most extreme values of vertical spin gaps of the singlet state structure $\Delta E_{\rm{s-t}/\!/s}$. 
In the case of cyclic carbenes, the steric constraint on the torsional angels can be the dominant factor of singlet spin state destabilization, e.\ g.\ we find a vinyl-carbene in a 7-cycle with $\Delta E_{\rm{s-t}/\!/s}>$0 kcal/mol (see Table\ SI-2).
Similarly, an aryl carbene with a positive value of $\Delta E_{\rm{s-t}/\!/s}$ (see inset structure in the Figure) illustrates how a fluorine hydrogen bond enforces a torsional angle on the carbene and a resulting $\pi$-interaction with an electrophilic aromatic system. 
\item Remarkably, a multitude of acyclic carbenes of group (c), namely amino-, carbonyl-, imino-, and vinyl-carbenes, are found among both the negative and positive extreme values of vertical spin gaps (see Tables SI-1 and SI-2).
For example, carbonyl-substituted methylidenes show negative spin gaps down to $\Delta E_{\rm{s-t}/\!/s}$ = -115.1 and positive spin gaps up to $E_{\rm{s-t}/\!/t}$ = 28.1.
This drastic change in the vertical spin gap is a showcase for how sensitive the spin gap is to $\pi$-interactions mediated via the torsional angles at the carbene center and has raised interest already for several decades.\cite{kim1980geometric}
Such a drastic change in chemical characteristic for a wide range of carbene classes and resulting vertical spin gaps has however not been reported before. 
\end{itemize}
\vspace{-0.7em}
%
%
\begin{figure}[t]
    \centering
    \includegraphics[width=1.075\linewidth]{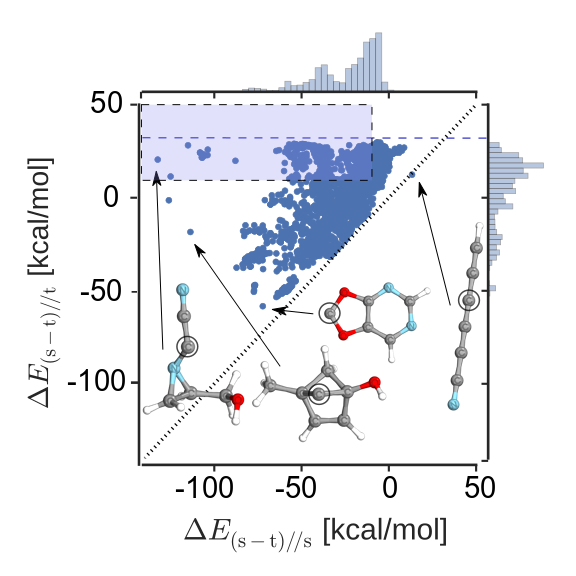}
    \caption{Spread of the singlet-triplet energy splittings (in kcal/mol) calculated with MRCISD-F12+Q/cc-pVDZ-F12 at the B3LYP geometry for the triplet state (y-axis) and CASSCF(2e,2o) geometry for the singlet state (x-axis). Distribution of the range of the singlet triplet gap energies among the carbene data set are given as joined plots.} 
    \label{fig:st_s_t}
\end{figure}
%
%
%
%
\begin{figure}[t]
    \centering
    \includegraphics[width=1.04\linewidth]{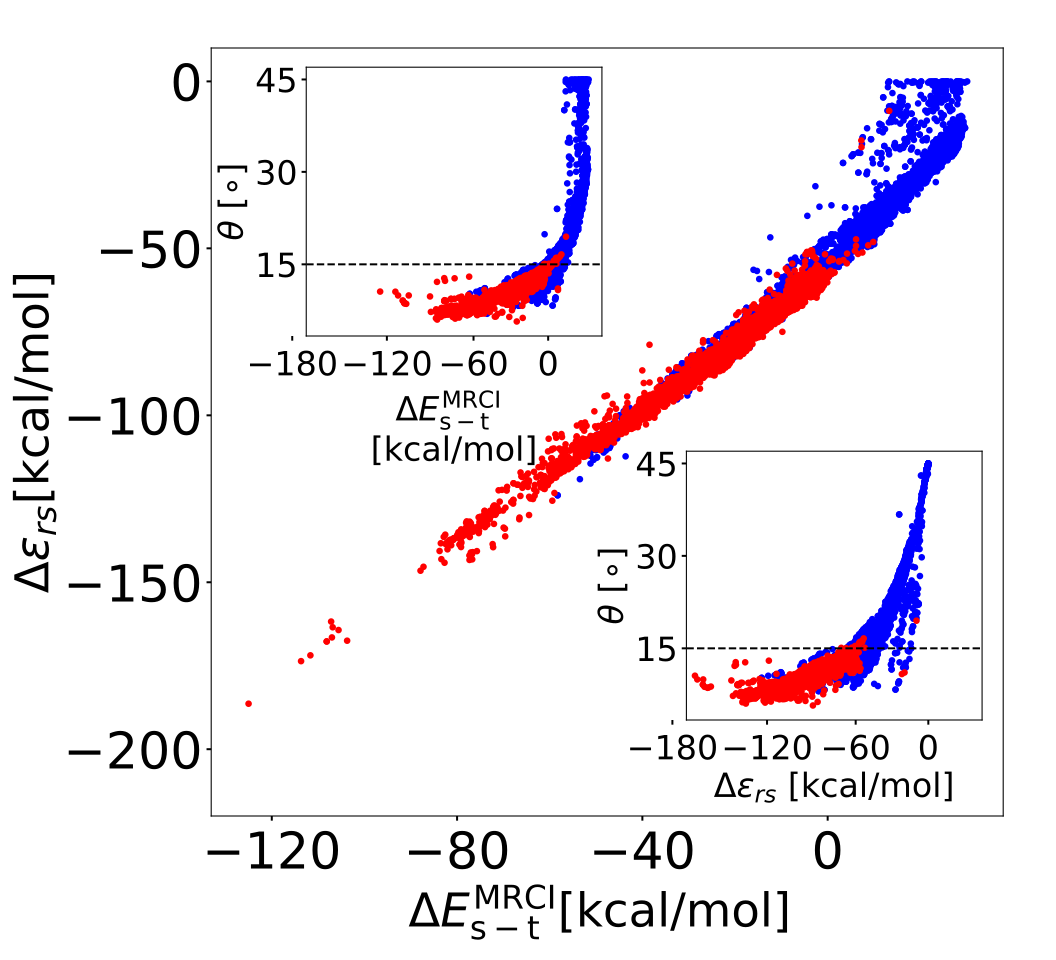}
    \caption{Correlation plot of the vertical singlet-triplet energy gap $E_{\mathrm{s-t}}$ (calculated using MRCISD+Q-F12/cc-pVDZ-F12) and the energy difference between the active orbitals at the carbene center  $\Delta \varepsilon_{rs}$ (calculated using SA-CASSCF/cc-pVDZ-F12) in kcal/mol. Correlation plots of the entanglement angle $\theta$ and $E_{\mathrm{s-t}}$ as well as $\theta$ and  $\Delta \varepsilon_{rs}$ are given as insets. Triplet and singlet state geometries are given in blue and red colour, respectively. }
    \label{fig:hl_ci_st}
\end{figure}
%


A scatter plot of the vertical spin gaps of the singlet and triplet state geometries is shown in Fig.\ \ref{fig:st_s_t}, illustrating the generally strong influence of the structural changes on the vertical spin gap.
Carbene centers with $sp$ carbons in both $\alpha$ positions can undergo maximum ``fourfold'' $\pi$-interactions by adopting themselves a $sp$ hybridization.
We find such a linear configuration in both singlet and triplet state structures for the dialkyne carbene HC$_6$N (see inset example in Fig.~\ref{fig:st_s_t}).
A particularity is that the singlet state is of $^1\Delta$ symmetry and therefore paramagnetic.
The triplet state has been studied experimentally\cite{gordon2000laboratory} and computationally,\cite{AOKI1994229} but only a stable ${}^1\Delta$ singlet state of the related linear diethynyl carbene has been studied computationally (at Hartree-Fock level of theory).\cite{cooper1988ab}
\indent The largest change in the vertical spin gap for the spin state specific geometries occurred in an amino-cyano-carbene (see inset Fig.\ \ref{fig:st_s_t}).
In the singlet state, strong $\pi$ donation from the geometrically constrained amino group leads to a large splitting ($\angle$ R-$\ddot{\rm{C}}$-R' =  111.5$^{\circ}$).
In the triplet state the carbene center approaches more of a $sp$ character, as two perpendicular $\pi$ systems are available for maximized ``threefold'' $\pi$-interaction ($\angle$ R-$\ddot{\rm{C}}$-R' = 134.9$^{\circ}$).
As another example of drastic changes in the vertical spin gap emerges a foiled carbene (see inset of the Figure) due to the strong through-space $\pi$-donation into the carbene $p$-orbital in the singlet state structure.\cite{gleiter1968stabilizing}
A notably large amount of carbenes in the QMspin data set are metastable with respect to inter-system-crossing by at least 10 kcal/mol, shown as the blue shaded area in Fig.\ \ref{fig:st_s_t}.
This means that in general a large number of carbenes can be accessible as metastable intermediates in both spin states. 
\\
\indent \textbf{A hard upper limit for $\Delta E_{\rm{s-t}}$.}
A very striking observation in Fig.\ \ref{fig:st_s_t} is the emergence of a hard upper limit of $\Delta E_{\rm{s-t}//t}$ of $\approx$30 kcal/mol on this diverse carbene chemical space. 
This finding is consistent with the fact that only
few computed adiabatic spin gaps of $\Delta E^{\mathrm{ad}}_{\rm{s-t}}>$20 kcal/mol are reported in the literature \cite{shainyan2013carbenes,nemirowski2007electronic,gronert2011stabilities} 
and that, to the best of our knowledge,
accurate predictions of spin gaps of $\Delta E^{\mathrm{ad}}_{\rm{s-t}}>$30 kcal/mol have
not yet been reported for non-metal $\alpha$-substituents.

CASSCF(2e,2o) yields a qualitatively correct description of the wave functions of both spin states with respect to MRCI (RMSD: 5.3 kcal/mol, max. error: 5.5 kcal/mol for the CASSCF vertical spin gap $\Delta E_{\rm{s-t}}^{\rm{CASSCF}}$).
An analysis of the CASSCF wave function is therefore well suited for identifying the significant physical interactions behind this phenomenon of a hard upper limit for the singlet-triplet vertical spin gap $\Delta E_{\rm{s-t}}$.
The correlation between the entanglement angle of the singlet state and $\Delta E_{\rm{s-t}}$ is depicted in the left inset of Fig.~\ref{fig:hl_ci_st} for the singlet state (red) and triplet state (blue) structures. 
Two different ``regimes'' can be clearly identified: (i) Regime of ``moderate'' carbene orbital entanglement with $\theta \le 15^{\circ}$ (resulting in $-150 \le \Delta E_{\rm{s-t}} \le$ 10 kcal/mol); (ii) Regime of strong carbene orbital entanglement with $\theta > 15^{\circ}$ (resulting in $5 \le \Delta E_{\rm{s-t}}\le 30 $ kcal/mol). 

The energy splitting $\Delta \varepsilon_{rs}$ on the carbene non-bonding orbitals $r$ and $s$ is a matter of long-standing scientific interest.\cite{gleiter1968stabilizing,andrada2015direct} 
Given that the active space of our state-averaged CASSCF calculations is verified to reproduce the nonbonding carbene orbitals we define the associated orbital energies $\varepsilon$ by borrowing the reasoning of Koopmans' Theorem\cite{koopmans1934zuordnung} as the state averaged  negative ionization energies of the closed shell singlet electron configuration associated with the orbital $r$ or $s$ and the triplet configuration.
This orbital energy can be computed as:
\begin{equation}
  \varepsilon_{r} = \varepsilon^{(1)}_r + \frac12 \big(J_{rr} + J_{rs} -K_{rs}\big) 
\end{equation}
with $\varepsilon^{(1)}$ containing the mean-field interaction with closed shell orbitals, the kinetic energy term, and electron--nuclei interaction. 
$J_{rs}$ is the Coulomb repulsion of two electrons in orbitals $r$ and $s$, $K_{rs}$ the corresponding exchange interaction in the triplet state.
The spin gap at CASSCF level of theory can then be expressed in terms of the entanglement angle $\theta$ as (see section 3 in the SI for a detailed derivation)
\begin{equation}
    \Delta E_{\rm{s-t}}^{\rm{CASSCF}} = \cos(2\theta)\ \Delta\varepsilon_{rs} + \big[ 1- \sin(2\theta)\big]K_{rs} + \Delta J_{rs}^+
    \label{eq:st_first}
\end{equation}
where $\Delta J_{rs}^+=\frac12 (J_{rr} + J_{ss}) -J_{rs}$ is the averaged difference of Coulomb interactions between electrons in the non-bonding orbitals for the singlet and triplet state configurations.
In eq.\ \ref{eq:st_first} it can be seen that a large energy splitting of the active space orbitals is a main driving force in the singlet state stabilization, whereas the minimization in Coulomb interaction within the active space, expressed by $\Delta J_{rs}^+>0$, is a main driving force in the triplet state stabilization.
The resonance energy stabilization via $K_{rs}$ is present in both states, but its magnitude has the prefactor $\cos(2\theta)$ in the singlet state.
The average contributions of $K_{rs}$ and $\Delta J_{rs}^+$ in our data set are $\approx$30 and $\approx$37 kcal/mol, respectively. 
Distribution plots of these quantities are shown in Fig. SI-4. 
$\Delta J_{rs}^+$ is rather narrowly distributed within a few kcal/mol.
The resonance interaction $K_{rs}$ depends more sensitively on the spatial overlap of the orbitals and therefore varies a bit stronger.

Taylor expansions for the moderately and strongly entangled regime can be obtained at the angles $\theta=0^\circ$ and $\theta=45^\circ$, respectively.
Using the variational condition $\frac{\partial \Delta E_{\rm{s-t}}}{\partial \theta}=0$ of CASSCF, the second order Taylor expansions yield approximate direct relationships of the spin gap and the orbital energy splitting as 
\begin{flalign}
   & \text{for} \ \theta \ge 25^{\circ}: \nonumber \\
   & \quad \Delta E_{\rm{s-t}}^{\rm{CASSCF}} \approx \frac{- \Delta \varepsilon_{rs}^2}{2K_{rs}} +   \Delta J_{rs}^+ \quad , \nonumber  \\ 
    & \text{for} \ \theta \le 15^{\circ}: \nonumber \\
    & \quad \Delta E_{\rm{s-t}}^{\rm{CASSCF}} \approx \Delta \varepsilon_{rs} + \Delta J_{rs}^+ + K_{rs}\left( 1+\frac{K_{rs}}{2\Delta \varepsilon_{rs}} \right) \ .  \label{eq:relgap}
\end{flalign}
Hence, in the moderately entangled regime, a linear relationship of $\Delta E_{\rm{s-t}}$ and $\Delta\varepsilon_{rs}$ would be approached with an offset given by $\Delta J_{rs}^++K_{rs}$. In the strongly entangled regime $\Delta E_{\rm{s-t}}$ would depend quadratically on the vanishing value of $\Delta\varepsilon_{rs}$, reaching an upper limit given by $\Delta J_{rs}^+$.
Similar relationships have been derived for valence configuration interaction calculations on singlet-triplet spin gaps of polyacenes.\cite{minami2013signature}
The relationships in eq.\ \ref{eq:relgap} can indeed be found for $\Delta E_{\rm{s-t}}$ in Fig.\ \ref{fig:st_s_t}: 
The mean signed deviation of CASSCF with respect to MRCI in our data set is -5.1 kcal/mol, arising from stronger dynamic electron correlation stabilization of the singlet state in comparison to the triplet state.
Subtracting these higher-order electron correlation effects, the contribution of $\Delta J_{rs}^+$ can consistently explain the upper limit.
In the moderately entangled regime the approximately linear relationship is also observed.
Hence, carbene spin gaps over the whole chemical space can be explained in an intuitive way. The upper limit for the vertical spin gap is inherent to the carbene characteristic of a divalent carbon atom with two non-bonding orbitals of mixed $s$ and $p$ character.




Gleiter and Hoffmann computed with extended H\"uckel calculations for a variety of carbenes in 1968 \cite{gleiter1968stabilizing} that an energy splitting of less than about 35 kcal/mol would lead to a triplet ground state.
In turn, values above about 50 kcal/mol would lead to a singlet ground state. 
In comparison, we find orbital energy splitting values of about 45 and 60 kcal/mol as the limit values for a triplet and singlet ground state, respectively.\\
\indent \textbf{Conjugation to strongly electrophilic $\pi$-systems.}
Among the molecules that have not been retained in the QMspin carbene data set because of strong delocalization of the carbene non-bonding orbitals were a significant number of carbenes conjugated to strongly electrophilic $\pi$-systems.
Aryl carbenes show stabilization of their triplet states by moderate delocalization 
of the diradical character.\cite{bourissou2000stable}
However, when coupled to $\pi$-systems with strong electrophilic character, it is generally reasoned that a complete mixing of the carbene $p$-orbital with the $\pi$-system occurs,\cite{hirai2009persistent} provoking the loss of the ``genuine'' carbene character.
In order to gain a qualitative insight, for a subset of such carbenes sorted out from the QMspin data set, we performed state-averaged CASSCF(4e,4o) calculations including the two lowest singlet and four lowest triplet states, followed by MRCI calculations of the individual states.
An example of such a molecular orbital diagram for the CASSCF active space is given in Fig.~SI-3, 
2D structures of a representative set along with their vertical spin gaps are shown in Table~SI-3.  
The active space natural orbitals yield in most cases a localized $sp^2$ orbital, as well as delocalized $\pi$-orbitals.
Strong orbital entanglement in the lowest singlet state of this extended active space then usually emerges in the form of diradical singlet configurations which would justify the attribution of these systems to the class of ``open shell singlet'' carbenes.\cite{borden2000interplay}
%
\begin{figure}
    \centering
    \includegraphics[width=1.03\linewidth]{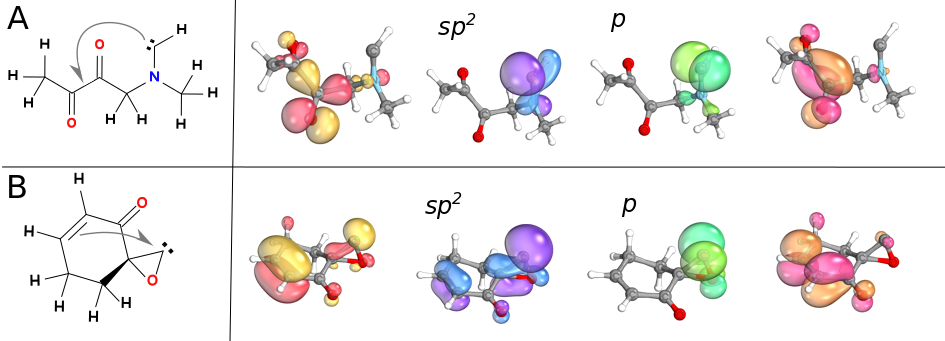}
    \caption{Molecular orbitals (MO) of the $sp^2$ and $p$ orbitals of the carbene center and the low lying HOMO and LUMO orbitals of interacting functional groups for molecules with considerable delocalization of the carbene non-bonding orbitals (molecules not retained in the QMspin set). Hyperconjugation is indicated by curly arrows in the 2D structure.}
    \label{fig:types_carbenes}
\end{figure}
\begin{figure}[t]
    \centering
    \includegraphics[scale=0.24]{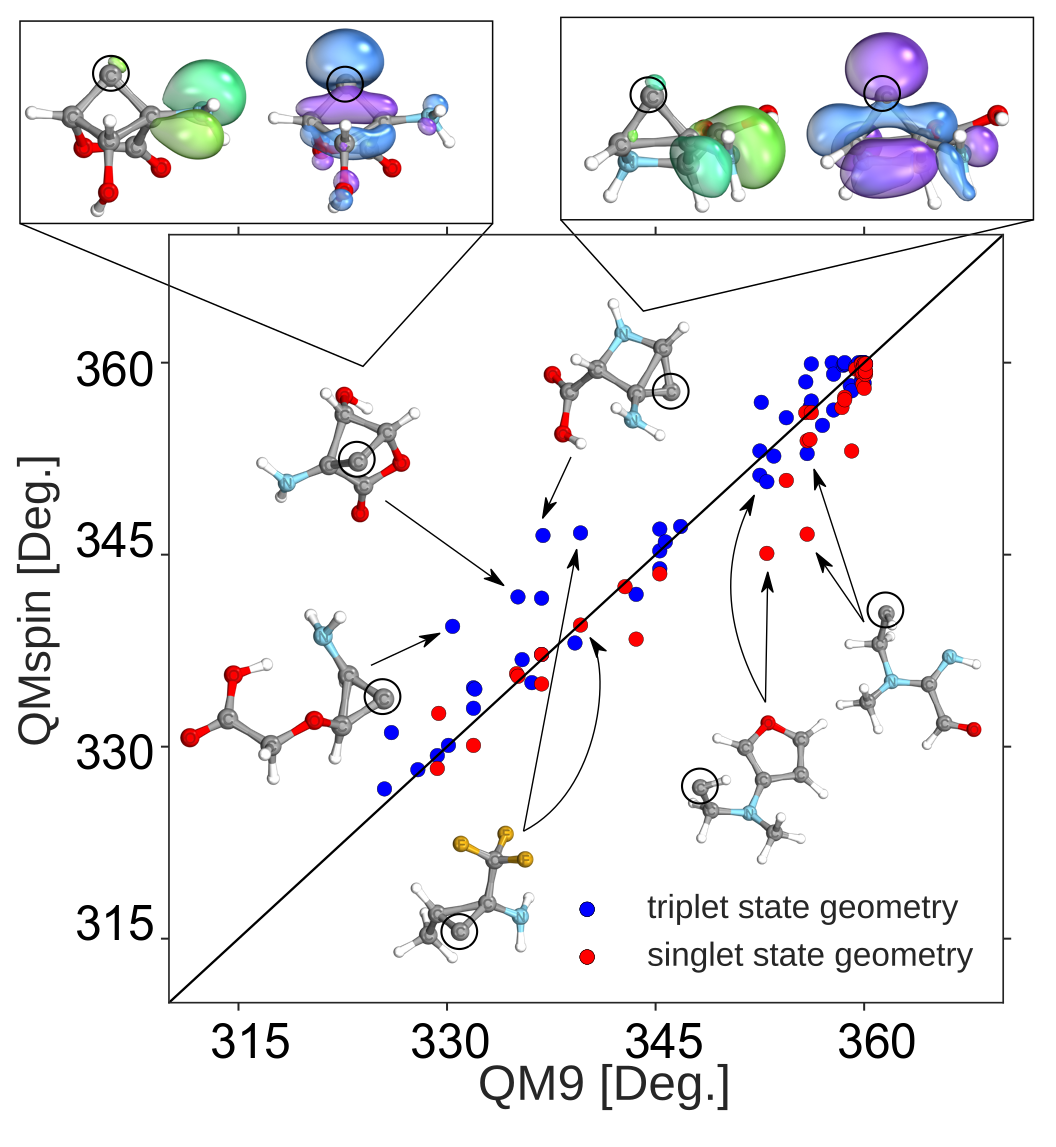}
    \caption{Sum of bond angles on the nitrogen atom in carbene molecules with an alkyl amino group at $\beta$-position to the carbene center for the singlet state and triplet state structure, as well as the corresponding QM9 molecule. Hyperconjugation is illustrated by the open shell restricted KS-B3LYP orbitals (the triplet state geometry optimization method) at the carbene center ($sp^2$ blue/purple), the nitrogen atom $p$-Orbitals are shown in green.}
    \label{fig:amino}
\end{figure}
\\
\indent \textbf{Hyperconjugation across saturated carbon.}
Oxy- and amino-carbenes may show high energy splittings of the non-bonding carbene orbitals and hence low-lying singlet states, especially if the carbene center is mesomerically ``isolated'' by a saturated carbon (or the $sp^3$ nitrogen of an amino-group).
Among systems that showed strong delocalization of the carbene non-bonding orbitals, we investigated some with a (4e,4o) active space, including the four lowest triplet states and the lowest singlet state in the  state-averaged CASSCF calculation. 
For two representative molecules {\bf A} and {\bf B}, the four natural orbitals of the active space are depicted in Figure \ref{fig:types_carbenes}.
Each of them represents a typical scenario for the breakdown of the carbene character:
\vspace{-0.7em}
\begin{itemize}
\setlength\itemsep{-0.3em}
\item {\bf A}: Strongly electrophilic $\pi$-systems with a rather low lying unoccupied orbital (e.\ g.\ 1,2-diketones) may be close in energy to the carbene $p$-orbital.
For several molecules with low-lying singlet states, this leads to the loss of a clear carbene character of the triplet state wave function in the sense that significant charge transfer from the carbene $p$-orbital to the strongly electrophilic group occurs in the triplet state structure.
{\bf A} localizes well the natural orbitals both on the carbene center and the $\pi$-systems with the extended (4e,4o) active space, yet the natural orbitals did not localize well with the (2e,2o) active space. 
The reason is that the lowest triplet state in the extended active state shows  non-negligible contributions from a configuration that involves charge transfer from the carbene center to the strongly electrophilic $\pi$-system.
Namely the charge transfer occurs into the LUMO of the diketone functional group (rightmost orbital of {\bf A} in Fig.\ \ref{fig:types_carbenes}, see Table SI-4 
for details on the electron configurations involved in the triplet state).
\vspace{-0.5em}
\item {\bf B}: The $sp^2$ carbene non-bonding orbital may fall in the energy range of $\pi$-orbitals of alkenyl groups.
The triplet ground state wave function shows indeed a strong contribution (16\%) from a configuration that expresses charge-transfer from the vinyl group to the $sp^2$ carbene orbital. 
The $sp^2$ carbene orbital of the triplet state optimized structure of molecule {\bf B} in Fig.~\ref{fig:types_carbenes} shows significant contributions on the vinyl group (14 \%) and the $sp^3$-carbon in $\alpha$-position to the carbene center (9 \%).
Hence, strong hyperconjugation involving the saturated ``$sp^5$'' carbon center\cite{bernett1967unified} of the 3-cycle is present in the triplet state.
\end{itemize}
\vspace{-0.7em}
These examples illustrate the limits of up to how much the carbene non-bonding orbitals can vary in energy in this chemical space  before the clear carbene electronic structure breaks down.

Finally, it is known that the carbene non-bonding orbitals can have interactions with $\pi$-orbitals in $\beta$-position beyond conjugation.\cite{lambert1997beta,johnson2008topological} 
In order to probe eventual similar unconventional effects of nitrogen containing groups in our data set, we investigated structural changes of acyclic amino groups in $\beta$-position of the carbene center, separated by a saturated carbon from the latter.
In Fig.\ \ref{fig:amino} the sum of bond angles of such nitrogen centers in the triplet and singlet optimized carbene structures is compared to the one of the corresponding closed-shell QM9 molecule. Thereby, the initial QM9 structure was reoptimized at the same level of theory as the triplet state structure (spin-restricted B3LYP/def2-TZVP).
A significant gain in planarity of the amine group in the triplet state structure is observed for carbenes derived from small cycles.
Three exemplary structures are shown as inset in the Figure, where the effect is strongest for cyclopropylidenes.\cite{jones1963cyclopropylidene}
We indeed find that the $\pi^{*}$-orbital of the N-C$_{\rm{cycle}}$ bond hyperconjugates with the carbene $sp^2$ non-bonding orbital and stabilizes the triplet state by spin density delocalization.
This singly occupied orbital, as well as the alkyl amine lone pair with mainly $p$-character, are shown as an inset for two of these structures in the Figure, computed as Kohn-Sham orbitals of the triplet state structure optimization method.
The active space orbitals of CASSCF(2e,2o) reproduce this picture.
To our best knowledge, this is the first reported example of such an amino-to-carbene interaction across a saturated carbon center. 
The destabilization of the saturated carbon center by strong ring constraints may strongly favor such hyperconjugation effects.


From Fig.\ \ref{fig:amino} it can also be seen that the singlet state structure of the carbenes,
on the other hand, shows rather lower planarity in the amino-group in $\beta$-position as compared to the closed shell structures, possibly induced by polarization effects (see two example molecules as inset in Fig.\ \ref{fig:amino}). \par
%
%
\bigskip
\textbf{Conclusion}
\\
\\
We present a higher--order multireference level of theory study of a large carbene chemical space with more than eight thousand distinct machine-generated molecular scaffolds. 
Accuracy of the data is ensured by a protocol that verifies the genuine carbene character of the electronic structure.
Unexpectedly, an overwhelming large part of the combinatorially possible carbene chemical space (almost 90\%) could be classified as genuine carbenes with a stable structure in both spin states, indicating the ubiquitous availability of carbenes as reaction intermediates in organic chemistry.

Chemical rules have been extracted as they emerge from the data, helping to reduce human selection bias based on preconceived assumptions and data-scarcity.
An inherent hard upper limit of about 30 kcal/mol in the singlet-triplet spin gap has been observed and an explanation based on the underlying physics of the carbene electronic structure along with numerical evidence has been provided.
Further unexpected insights on the chemical space include large vertical and adiabatic spin gap ranges for many carbene classes ($>$60 and $>$100 kcal/mol, respectively), as well as the  prediction of a linear singlet state carbene in the paramagnetic $^1\Delta$ state at multi-reference level of theory.
The delicate role of conjugation and hyperconjugation over the chemical space has been elucidated, as these offer energetic stabilization but may ultimately lead to the breakdown of the carbene character if too dominant.
To the best of our knowledge, this is the first example of an approach that seeks to probe the inherent compositional limitations of a large chemical space of machine-generated scaffolds with systematic application of higher-order multi-reference level of theory.

Understanding and mapping composition and atomic configuration to electronic structure for a chemical space of free carbenes opens the door towards libraries of energetically accessible intermediates which are crucial for the construction of organic chemistry reaction networks.
Furthermore, we provide the largest set published so far for accurate spin gaps of molecules with strongly correlated electrons, ideal for benchmarking computationally more efficient but less accurate quantum chemical methods. Finally, the data could also serve the training, testing, and application of machine learning models. 

\bigskip
\textbf{Methods}
\\
\\
\textbf{Data set generation}.
A carbene chemical space of $\approx$8000 molecules was created from $\approx$4000 randomly selected molecular geometries out of the $\approx$130k molecules of the QM9 \cite{ramakrishnan2014quantum} data set. The created data set is named ``QMspin'' here-within.
Two hydrogen atoms were abstracted from all applicable saturated carbon centers in the original molecules using the OpenEye Toolkit. \cite{OpenEye}
The triplet state carbene geometry was optimized using open-shell restricted Kohn-Sham B3LYP\cite{becke1988density,lee1988development} with the def2-TZVP\cite{weigend2005balanced} molecular orbitals basis and the def2-TZVPP density fitting basis.
The carbene singlet state geometries were optimized with the complete active space self consistent field (CASSCF) method\cite{werner1985jcp,werner:2019,Busch1991JCP} implemented in Molpro\cite{MOLPRO_brief} with a two electrons in two orbitals [(2e,2o)] active space using the cc-pVDZ-F12\cite{peterson2008systematically} (``VDZ-F12'') molecular orbital basis and the aug-cc-pVTZ\cite{kirk:vnzf1208} density fitting basis.
The singlet-state geometry optimizations were started from the triplet state structures.
Since the singlet state geometry optimizations were considerably more expensive in terms of computational cost, they were done for $\approx$60\% of the triplet state structures.
The data set therefore contains $\approx$13k carbene structures of $\approx$8000 distinct scaffolds.
Molecules that showed bond breaking during the geometry optimizations have been sorted out ($\approx$8\%).\cite{bondbreak}
To verify that the geometries optimized at singlet and triplet state converged to the same conformational local minima, we computed the RMSD of the changes in atomic positions between them. 
For the few cases where RMSD $> 1 $ {\AA} it was verified that the large RMSD was due to changes in bond and torsion angle on the carbene center itself.
The latter arises solely as a direct consequence of the change in spin state. \\ \indent 
The multireference SCF spin gaps for the optimized geometries have been computed with the state-averaged (SA) CASSCF(2e,2o) method.
Based on this reference wave function, internally contracted explicitly correlated multi-reference configuration interaction singles and doubles with the quadruples Davidson correction (MRCISD+Q-F12)\cite{werner1988jcp,knowles1992internally,Knowles1988CPL,shiozaki2011explicitly,DavidsonCorr} singlet and triplet state energies were computed using the same basis set as for the CASSCF computations.
The application of the F12 correction generally enables obtaining results at MRCISD+Q/quadruple zeta basis quality already at the MRCISD+Q-F12/VDZ-F12 level of theory.\cite{shiozaki2011explicitly} 
The complementary auxiliary basis set (CABS) singles correction\cite{knizia2008explicitly} is included in the MRCISD+Q-F12 results.
For the spin gap in methylene, the basis set convergence of MRCISD+Q-F12/VDZ-F12 was indeed found to be within chemical accuracy to the basis set limit.\cite{shiozaki2011explicitly}
For further detailed discussion on the suitability of our methodological approach see Section 1 in the SI.
Orbital visualization has been realized with IboView.\cite{IboView}
 \\
\noindent \textbf{Identification of the carbene character.}
It was verified that the computed SA-CASSCF(2e,2o) wave function corresponds to the electronic structure of a carbene.
To this end, we localized the closed and active orbital spaces by computing intrinsic bond orbitals (IBOs).\cite{knizia2013intrinsic}
Subsequently, we used these to derive bonding valencies based on the closed-shell orbitals for all carbon atoms in the molecule. 
A closed shell orbital was attributed to all carbon atoms with a contribution of $\ge 20$\% to the bond. The IBO orbital localization procedure was chosen with exponent ``4'' such that non-polar aromatic compounds are described by bonds mainly located at two centers.\cite{knizia2013intrinsic}
In this way, moderate (hyper-)conjugation effects in the closed shell orbitals do not lead to an erroneous bond valency overcount for the carbon atoms.
For active space orbitals, a contribution of $\ge 30$\% from the carbene center was required and it was verified that no other carbon atom bears a contribution of $\ge 30$\%.
A correct carbene description is then given by the presence of only one divalent carbon atom, all other carbon atoms being tetravalent according to the connectivity established with IBOs.
This was found for the vast majority of the initial triplet state structures ($\approx$ 95\%) and almost all of the initial singlet state structures ($\approx 99.5\%$).
Among the triplet state optimized structures retracted from the QMspin data set (``non-genuine'' carbenes), we investigated a subset with a larger CASSCF active space (see Results and Discussion section for details).
For the molecules retained in the QMspin data set it was also verified that the wave function of the method used for geometry optimization [restricted open-shell B3LYP or singlet state CASSCF(2e,2o)] corresponded to a carbene in the sense defined above.
Furthermore, it was verified that the weight of the reference configurations in the normalized MRCI wave function was $>0.75$. 
This left us with 8062 triplet and 5021 singlet state optimized structures in the QMspin data set. 

\bigskip

\textbf{Supporting Information} \\
Additional figures and tables are shown and additional aspects on the choice of the used quantum chemistry methods are discussed. The equations for the approximate relationship of carbene non-bonding orbitals and vertical singlet-triplet spin gaps are derived in detail.
The data of the underlying quantum chemistry calculations (structures, energies at different levels of theory, orbital partial charges, and further calculation parameters) will be made available on www.materialscloud.org.

\bigskip

\textbf{Acknowledgments.} \\
We acknowledge support by the European Research Council (ERC-CoG grant QML) as well as by the Swiss National Science foundation (No.~PP00P2\_138932, 407540\_167186 NFP 75 Big Data, 200021\_175747, NCCR MARVEL).
Some calculations were performed at sciCORE (http://scicore.unibas.ch/) scientific computing core facility at University of Basel.

\bibliographystyle{ieeetr}
\bibliography{references_new}  

\begin{thebibliography}{10}

\bibitem{Woodworth1959JACS}
R.~C. Woodworth and P.~S. Skell, ``{Methylene, CH2. Stereospecific Reaction
  with cis- and trans-2-Butene},'' {\em J. Am. Chem. Soc.}, vol.~81, no.~13,
  pp.~3383--3386, 1959.

\bibitem{csizmadia1968mechanism}
I.~G. Csizmadia, J.~Font, and O.~P. Strausz, ``{Mechanism of the Wolff
  rearrangement},'' {\em J. Am. Chem. Soc.}, vol.~90, no.~26, pp.~7360--7361,
  1968.

\bibitem{Chu2016JACS}
J.~Chu, D.~Munz, R.~Jazzar, M.~Melaimi, and G.~Bertrand, ``{Synthesis of
  Hemilabile Cyclic (Alkyl)(amino)carbenes (CAACs) and Applications in
  Organometallic Chemistry},'' {\em J. Am. Chem. Soc.}, vol.~138, no.~25,
  pp.~7884--7887, 2016.

\bibitem{arduengo1991stable}
A.~J. Arduengo~III, R.~L. Harlow, and M.~Kline, ``{A stable crystalline
  carbene},'' {\em J. Am. Chem. Soc.}, vol.~113, no.~1, pp.~361--363, 1991.

\bibitem{arduengo19981}
A.~J. Arduengo~III, J.~R. Goerlich, R.~Krafczyk, and W.~J. Marshall, ``{1, 3,
  4, 5--Tetraphenylimidazol--2-ylidene: The Realization of Wanzlick's Dream},''
  {\em Angew. Chem.}, vol.~37, no.~13-14, pp.~1963--1965, 1998.

\bibitem{igau1988analogous}
A.~Igau, H.~Grutzmacher, A.~Baceiredo, and G.~Bertrand, ``{Analogous
  $\alpha$,$\alpha'$-bis-carbenoid, triply bonded species: synthesis of a
  stable $\lambda^3$-phosphino carbene-$\lambda^5$-phosphaacetylene},'' {\em
  Journal of the American Chemical Society}, vol.~110, no.~19, pp.~6463--6466,
  1988.

\bibitem{sander1993carbenes}
W.~Sander, G.~Bucher, and S.~Wierlacher, ``{Carbenes in matrixes: spectroscopy,
  structure, and reactivity},'' {\em Chem. Rev.}, vol.~93, no.~4,
  pp.~1583--1621, 1993.

\bibitem{modarelli1991interception}
D.~A. Modarelli and M.~S. Platz, ``{Interception of dimethylcarbene with
  pyridine: a laser flash photolysis study},'' {\em J. Am. Chem. Soc.},
  vol.~113, no.~23, pp.~8985--8986, 1991.

\bibitem{droege2010angew}
T.~Dr{\"o}ge and F.~Glorius, ``The measure of all rings—n-heterocyclic
  carbenes,'' {\em Angew. Chem. Int. Ed.}, vol.~49, no.~39, pp.~6940--6952,
  2010.

\bibitem{shavitt1985geometry}
I.~Shavitt, ``{Geometry and singlet-triplet energy gap in methylene: {A}
  critical review of experimental and theoretical determinations},'' {\em
  Tetrahedron}, vol.~41, no.~8, pp.~1531--1542, 1985.

\bibitem{gerbig2013computational}
D.~Gerbig and D.~Ley, ``{Computational methods for contemporary carbene
  chemistry},'' {\em WIREs Comput. Mol. Sci.}, vol.~3, no.~3, pp.~242--272,
  2013.

\bibitem{gronert2011stabilities}
S.~Gronert, J.~R. Keeffe, and R.~A. More~O'Ferrall, ``{Stabilities of carbenes:
  {I}ndependent measures for singlets and triplets},'' {\em J. Am. Chem. Soc.},
  vol.~133, no.~10, pp.~3381--3389, 2011.

\bibitem{mieusset2008carbene}
J.-L. Mieusset and U.~H. Brinker, ``{The carbene reactivity surface: a
  classification},'' {\em J. Org. Chem.}, vol.~73, no.~4, pp.~1553--1558, 2008.

\bibitem{vasiliu2017characterization}
M.~Vasiliu, K.~A. Peterson, A.~J. Arduengo, and D.~A. Dixon,
  ``{Characterization of Carbenes via Hydrogenation Energies, Stability, and
  Reactivity: What's in a Name?},'' {\em Chem. Eur. J.}, vol.~23, no.~69,
  pp.~17556--17565, 2017.

\bibitem{shainyan2013carbenes}
B.~A. Shainyan, A.~V. Kuzmin, and M.~Y. Moskalik, ``{Carbenes and nitrenes. An
  overview},'' {\em Comput. Theor. Chem.}, vol.~1006, pp.~52--61, 2013.

\bibitem{ramakrishnan2014quantum}
R.~Ramakrishnan, P.~O. Dral, M.~Rupp, and O.~A. Von~Lilienfeld, ``{Quantum
  chemistry structures and properties of 134 kilo molecules},'' {\em Sci.
  Data}, vol.~1, p.~140022, 2014.

\bibitem{barluenga20088+}
J.~Barluenga, J.~Garc{\'\i}a-Rodr{\'\i}guez, S.~Martinez, A.~L. Suarez-Sobrino,
  and M.~Tomas, ``{[8+ 2] and [8+ 3] Cyclization Reactions of Alkenyl Carbenes
  and 8--Azaheptafulvenes: Direct Access to Tetrahydro--1--azaazulene and
  Cyclohepta [b] pyridinone Derivatives},'' {\em Chem. Asian J.}, vol.~3,
  no.~4, pp.~767--775, 2008.

\bibitem{seburg2009structure}
R.~A. Seburg, E.~V. Patterson, and R.~J. McMahon, ``{Structure of Triplet
  Propynylidene ({HCCCH}) as Probed by {IR}, {UV}/vis, and {EPR} Spectroscopy
  of Isotopomers},'' {\em J. Am. Chem. Soc.}, vol.~131, no.~26, pp.~9442--9455,
  2009.

\bibitem{lavallo2004synthesis}
V.~Lavallo, J.~Mafhouz, Y.~Canac, B.~Donnadieu, W.~W. Schoeller, and
  G.~Bertrand, ``{Synthesis, reactivity, and ligand properties of a stable
  alkyl carbene},'' {\em J. Am. Chem. Soc.}, vol.~126, no.~28, pp.~8670--8671,
  2004.

\bibitem{cattoen2004amino}
X.~Catto{\"e}n, H.~Gornitzka, D.~Bourissou, and G.~Bertrand,
  ``{Amino-aryl-carbenes: {A}lternative ligands for transition metals?},'' {\em
  J. Am. Chem. Soc.}, vol.~126, no.~5, pp.~1342--1343, 2004.

\bibitem{freeman2006singlet}
F.~Freeman and M.~Gomarooni, ``{Singlet--triplet gaps and insertion reactions
  of aminocyanocarbenes: {A} computational study of hydrogen cyanide covalent
  dimers},'' {\em J. Chem. Theory Comput.}, vol.~106, no.~11, pp.~2379--2389,
  2006.

\bibitem{bourissou2000stable}
D.~Bourissou, O.~Guerret, F.~P. Gabbai, and G.~Bertrand, ``{Stable carbenes},''
  {\em Chem. Rev.}, vol.~100, no.~1, pp.~39--92, 2000.

\bibitem{tomioka1997persistent}
H.~Tomioka, ``{Persistent triplet carbenes},'' {\em Acc. Chem. Res.}, vol.~30,
  no.~8, pp.~315--321, 1997.

\bibitem{hirai2009persistent}
K.~Hirai, T.~Itoh, and H.~Tomioka, ``{Persistent triplet carbenes},'' {\em
  Chem. Rev.}, vol.~109, no.~8, pp.~3275--3332, 2009.

\bibitem{toscano1994carboethoxycarbene}
J.~P. Toscano, M.~S. Platz, V.~Nikolaev, and V.~Popic, ``{Carboethoxycarbene. A
  laser flash photolysis study},'' {\em J. Am. Chem. Soc.}, vol.~116, no.~18,
  pp.~8146--8151, 1994.

\bibitem{scott2001singlet}
A.~P. Scott, M.~S. Platz, and L.~Radom, ``{Singlet- Triplet Splittings and
  Barriers to Wolff Rearrangement for Carbonyl Carbenes},'' {\em J. Am. Chem.
  Soc.}, vol.~123, no.~25, pp.~6069--6076, 2001.

\bibitem{kim1980geometric}
K.~S. Kim and H.~F. Schaefer~III, ``{Geometric isomerism in triplet carbenes:
  carbohydroxycarbene},'' {\em J. Am. Chem. Soc.}, vol.~102, no.~16,
  pp.~5389--5390, 1980.

\bibitem{w1988}
D.~Weininger, ``{{SMILES}, a chemical language and information system.
  1.~{I}ntroduction to methodology and encoding rules},'' {\em J. Chem. Inf.
  Comput. Sci.}, vol.~28, no.~1, pp.~31--36, 1988.

\bibitem{soleilhavoup2014cyclic}
M.~Soleilhavoup and G.~Bertrand, ``{Cyclic (alkyl)(amino) carbenes (CAACs):
  Stable carbenes on the rise},'' {\em Acc. Chem. Res.}, vol.~48, no.~2,
  pp.~256--266, 2014.

\bibitem{martin2011stable}
D.~Martin, M.~Soleilhavoup, and G.~Bertrand, ``{Stable singlet carbenes as
  mimics for transition metal centers},'' {\em Chem. Sci.}, vol.~2, no.~3,
  pp.~389--399, 2011.

\bibitem{tonner2007bonding}
R.~Tonner, G.~Heydenrych, and G.~Frenking, ``{Bonding Analysis of
  {N}--Heterocyclic Carbene Tautomers and Phosphine Ligands in
  {T}ransition--{M}etal Complexes: {A} Theoretical Study},'' {\em Chem. Asian
  J.}, vol.~2, no.~12, pp.~1555--1567, 2007.

\bibitem{minami2013signature}
T.~Minami, S.~Ito, and M.~Nakano, ``{Signature of Singlet Open-Shell Character
  on the Optically Allowed Singlet Excitation Energy and Singlet--Triplet
  Energy Gap},'' {\em J. Phys. Chem. A}, vol.~117, no.~9, pp.~2000--2006, 2013.

\bibitem{gordon2000laboratory}
V.~D. Gordon, M.~McCarthy, A.~Apponi, and P.~Thaddeus, ``{Laboratory detection
  of HC6N, a carbon chain with a triplet electronic ground state},'' {\em
  Astrophys. J.}, vol.~540, no.~1, p.~286, 2000.

\bibitem{AOKI1994229}
K.~Aoki and S.~Ikuta, ``{The singlet with a C3 ring: the probable candidate of
  HC6N and C7H2},'' {\em J. Mol. Struct. THEOCHEM}, vol.~310, pp.~229 -- 238,
  1994.

\bibitem{cooper1988ab}
D.~L. Cooper and S.~C. Murphy, ``{Ab Initio Geometries for C$_{2n+1}$H,
  C$_{2n+1}$H$^+$, and C_$2n+1$H$_2$ Species for N= 1, 2, 3},'' {\em Astrophys.
  J.}, vol.~333, p.~482, 1988.

\bibitem{gleiter1968stabilizing}
R.~Gleiter and R.~Hoffmann, ``{Stabilizing a singlet methylene},'' {\em J. Am.
  Chem. Soc.}, vol.~90, no.~20, pp.~5457--5460, 1968.

\bibitem{nemirowski2007electronic}
A.~Nemirowski and P.~R. Schreiner, ``{Electronic stabilization of ground state
  triplet carbenes},'' {\em J. Org. Chem.}, vol.~72, no.~25, pp.~9533--9540,
  2007.

\bibitem{andrada2015direct}
D.~M. Andrada, N.~Holzmann, T.~Hamadi, and G.~Frenking, ``{Direct estimate of
  the internal $\pi$-donation to the carbene centre within N-heterocyclic
  carbenes and related molecules},'' {\em Beilstein J. Org. Chem.}, vol.~11,
  no.~1, pp.~2727--2736, 2015.

\bibitem{koopmans1934zuordnung}
T.~Koopmans, ``{{\"U}ber die Zuordnung von Wellenfunktionen und Eigenwerten zu
  den einzelnen Elektronen eines Atoms},'' {\em Physica}, vol.~1, no.~1-6,
  pp.~104--113, 1934.

\bibitem{borden2000interplay}
W.~T. Borden, N.~P. Gritsan, C.~M. Hadad, W.~L. Karney, C.~R. Kemnitz, and
  M.~S. Platz, ``{The interplay of theory and experiment in the study of
  phenylnitrene},'' {\em Acc. Chem. Res.}, vol.~33, no.~11, pp.~765--771, 2000.

\bibitem{bernett1967unified}
W.~A. Bernett, ``{A unified theory of bonding for cyclopropanes},'' {\em J.
  Chem. Educ.}, vol.~44, no.~1, p.~17, 1967.

\bibitem{lambert1997beta}
J.~B. Lambert and X.~Liu, ``{The $\beta$-Heteroatom Effect on Carbenes},'' {\em
  Tetrahedron}, vol.~53, no.~29, pp.~9989--9996, 1997.

\bibitem{johnson2008topological}
L.~E. Johnson and D.~B. DuPr{\'e}, ``{Topological and orbital-based mechanisms
  of the electronic stabilization of bis (diisopropylamino)
  cyclopropenylidene},'' {\em J. Phys. Chem. A}, vol.~112, no.~32,
  pp.~7448--7454, 2008.

\bibitem{jones1963cyclopropylidene}
W.~Jones, M.~H. Grasley, and W.~S. Brey, ``{The Cyclopropylidene 1: Generation
  and Reactions},'' {\em J. Am. Chem. Soc.}, vol.~85, no.~18, pp.~2754--2759,
  1963.

\bibitem{OpenEye}
``{OpenEye Toolkits:} {O}peneye {S}cientific {S}oftware, {S}anta {F}e, {NM}..''
  \url{http://www.eyesopen.com}.

\bibitem{becke1988density}
A.~D. Becke, ``{Density-functional exchange-energy approximation with correct
  asymptotic behavior},'' {\em Phys. Rev. A}, vol.~38, no.~6, pp.~3098--3100,
  1988.

\bibitem{lee1988development}
C.~Lee, W.~Yang, and R.~G. Parr, ``{Development of the Colle-Salvetti
  correlation-energy formula into a functional of the electron density},'' {\em
  Phys. Rev. B}, vol.~37, no.~2, pp.~785--789, 1988.

\bibitem{weigend2005balanced}
F.~Weigend and R.~Ahlrichs, ``{Balanced basis sets of split valence, triple
  zeta valence and quadruple zeta valence quality for {H} to {R}n: {D}esign and
  assessment of accuracy},'' {\em Phys. Chem. Chem. Phys.}, vol.~7, no.~18,
  pp.~3297--3305, 2005.

\bibitem{werner1985jcp}
H.-J. Werner and P.~J. Knowles, ``{A second order multiconfiguration SCF
  procedure with optimum convergence},'' {\em J. Chem. Phys.}, vol.~82, no.~11,
  pp.~5053--5063, 1985.

\bibitem{werner:2019}
D.~A. Kreplin, P.~J. Knowles, and H.-J. Werner, ``Second-order mcscf
  optimization revisited. i. improved algorithms for fast and robust
  second-order casscf convergence,'' {\em J. Chem. Phys.}, vol.~150, no.~19,
  2019.

\bibitem{Busch1991JCP}
T.~Busch, A.~D. Esposti, and H.~Werner, ``{Analytical energy gradients for
  multiconfiguration self--consistent field wave functions with frozen core
  orbitals},'' {\em J. Chem. Phys.}, vol.~94, no.~10, pp.~6708--6715, 1991.

\bibitem{MOLPRO_brief}
H.-J. Werner, P.~J. Knowles, G.~Knizia, F.~R. Manby, M.~{Sch\"{u}tz}, {\em
  et~al.}, ``{MOLPRO, version 2018.1, a package of ab initio programs},'' 2018.

\bibitem{peterson2008systematically}
K.~A. Peterson, T.~B. Adler, and H.-J. Werner, ``{Systematically convergent
  basis sets for explicitly correlated wavefunctions: {T}he atoms H, He, B--Ne,
  and Al--Ar},'' {\em J. Chem. Phys.}, vol.~128, no.~8, p.~084102, 2008.

\bibitem{kirk:vnzf1208}
K.~A. Peterson, T.~B. Adler, and H.-J. Werner, ``{Systematically convergent
  basis sets for explicitly correlated wavefunctions: The atoms H, He, B-Ne and
  Al-Ar},'' {\em J. Chem. Phys.}, vol.~128, p.~084102, 2008.

\bibitem{bondbreak}
The protein data bank (PDB)\cite{pdb} format has been used for this task.

\bibitem{werner1988jcp}
H.-J. Werner and P.~J. Knowles, ``{An efficient internally contracted
  multiconfiguration-reference configuration interaction method},'' {\em J.
  Chem. Phys.}, vol.~89, no.~9, pp.~5803--5814, 1988.

\bibitem{knowles1992internally}
P.~J. Knowles and H.-J. Werner, ``{Internally contracted
  multiconfiguration--reference configuration interaction calculations for
  excited states},'' {\em Theor. Chim. Acta}, vol.~84, no.~1-2, pp.~95--103,
  1992.

\bibitem{Knowles1988CPL}
P.~J. Knowles and H.-J. Werner, ``An efficient method for the evaluation of
  coupling coefficients in configuration interaction calculations,'' {\em Chem.
  Phys. Lett.}, vol.~145, no.~6, pp.~514 -- 522, 1988.

\bibitem{shiozaki2011explicitly}
T.~Shiozaki, G.~Knizia, and H.-J. Werner, ``{Explicitly correlated
  multireference configuration interaction: {MRCI-F12}},'' {\em J. Chem.
  Phys.}, vol.~134, no.~3, p.~034113, 2011.

\bibitem{DavidsonCorr}
The Davidson correction of the relaxed reference coefficients has been used.

\bibitem{knizia2008explicitly}
G.~Knizia and H.-J. Werner, ``{Explicitly correlated {RMP2} for high--spin
  open-shell reference states},'' {\em J. Chem. Phys.}, vol.~128, no.~15,
  p.~154103, 2008.

\bibitem{IboView}
IboView Orbital visualization program, G. Knizia
  \texttt{https://www.iboview.org}, Accessed: 09.04.2020.

\bibitem{knizia2013intrinsic}
G.~Knizia, ``{Intrinsic atomic orbitals: An unbiased bridge between quantum
  theory and chemical concepts},'' {\em J. Chem. Theory Comput.}, vol.~9,
  no.~11, pp.~4834--4843, 2013.

\bibitem{pdb}
H.M. Berman, J. Westbrook, Z. Feng, G. Gilliland, T.N. Bhat, H. Weissig, I.N.
  Shindyalov, P.E. Bourne, "The Protein Data Bank", \textit{Nucleic Acids
  Res.}, vol. 28, no. pp. 235--242, 2000, https://rcsb.org.

\end{thebibliography}


\providecommand{\latin}[1]{#1}
\makeatletter
\providecommand{\doi}
  {\begingroup\let\do\@makeother\dospecials
  \catcode`\{=1 \catcode`\}=2 \doi@aux}
\providecommand{\doi@aux}[1]{\endgroup\texttt{#1}}
\makeatother
\providecommand*\mcitethebibliography{\thebibliography}
\csname @ifundefined\endcsname{endmcitethebibliography}
  {\let\endmcitethebibliography\endthebibliography}{}
\begin{mcitethebibliography}{11}
\providecommand*\natexlab[1]{#1}
\providecommand*\mciteSetBstSublistMode[1]{}
\providecommand*\mciteSetBstMaxWidthForm[2]{}
\providecommand*\mciteBstWouldAddEndPuncttrue
  {\def\EndOfBibitem{\unskip.}}
\providecommand*\mciteBstWouldAddEndPunctfalse
  {\let\EndOfBibitem\relax}
\providecommand*\mciteSetBstMidEndSepPunct[3]{}
\providecommand*\mciteSetBstSublistLabelBeginEnd[3]{}
\providecommand*\EndOfBibitem{}
\mciteSetBstSublistMode{f}
\mciteSetBstMaxWidthForm{subitem}{(\alph{mcitesubitemcount})}
\mciteSetBstSublistLabelBeginEnd
  {\mcitemaxwidthsubitemform\space}
  {\relax}
  {\relax}

\bibitem[Feller \latin{et~al.}(1982)Feller, McMurchie, Borden, and
  Davidson]{feller1982theoretical}
Feller,~D.; McMurchie,~L.~E.; Borden,~W.~T.; Davidson,~E.~R. A theoretical
  determination of the electron affinity of methylene. \emph{J. Chem. Phys.}
  \textbf{1982}, \emph{77}, 6134--6143\relax
\mciteBstWouldAddEndPuncttrue
\mciteSetBstMidEndSepPunct{\mcitedefaultmidpunct}
{\mcitedefaultendpunct}{\mcitedefaultseppunct}\relax
\EndOfBibitem
\bibitem[Shavitt(1985)]{shavitt1985geometry}
Shavitt,~I. Geometry and singlet-triplet energy gap in methylene: {A} critical
  review of experimental and theoretical determinations. \emph{Tetrahedron}
  \textbf{1985}, \emph{41}, 1531--1542\relax
\mciteBstWouldAddEndPuncttrue
\mciteSetBstMidEndSepPunct{\mcitedefaultmidpunct}
{\mcitedefaultendpunct}{\mcitedefaultseppunct}\relax
\EndOfBibitem
\bibitem[Irikura \latin{et~al.}(1992)Irikura, Goddard~III, and
  Beauchamp]{irikura1992singlet}
Irikura,~K.~K.; Goddard~III,~W.; Beauchamp,~J. {Singlet-triplet gaps in
  substituted carbenes CXY (X, Y= H, fluoro, chloro, bromo, iodo, silyl)}.
  \emph{J. Am. Chem. Soc.} \textbf{1992}, \emph{114}, 48--51\relax
\mciteBstWouldAddEndPuncttrue
\mciteSetBstMidEndSepPunct{\mcitedefaultmidpunct}
{\mcitedefaultendpunct}{\mcitedefaultseppunct}\relax
\EndOfBibitem
\bibitem[Bauschlicher~Jr \latin{et~al.}(1987)Bauschlicher~Jr, Langhoff, and
  Taylor]{bauschlicher19871}
Bauschlicher~Jr,~C.~W.; Langhoff,~S.~R.; Taylor,~P.~R. On the
  ${}^{1}$A$_{1}$--${}^{3}$B$_{1}$ separation in CH$_2$ and SiH$_2$. \emph{J.
  Chem. Phys.} \textbf{1987}, \emph{87}, 387--391\relax
\mciteBstWouldAddEndPuncttrue
\mciteSetBstMidEndSepPunct{\mcitedefaultmidpunct}
{\mcitedefaultendpunct}{\mcitedefaultseppunct}\relax
\EndOfBibitem
\bibitem[Langhoff and Davidson(1974)Langhoff, and
  Davidson]{langhoff1974configuration}
Langhoff,~S.~R.; Davidson,~E.~R. Configuration interaction calculations on the
  nitrogen molecule. \emph{Int. J. Quantum Chem.} \textbf{1974}, \emph{8},
  61--72\relax
\mciteBstWouldAddEndPuncttrue
\mciteSetBstMidEndSepPunct{\mcitedefaultmidpunct}
{\mcitedefaultendpunct}{\mcitedefaultseppunct}\relax
\EndOfBibitem
\bibitem[Yamaguchi \latin{et~al.}(1993)Yamaguchi, Okumura, Mori, Maki, Takada,
  Noro, and Tanaka]{yamaguchi1993comparison}
Yamaguchi,~K.; Okumura,~M.; Mori,~W.; Maki,~J.; Takada,~K.; Noro,~T.;
  Tanaka,~K. Comparison between spin restricted and unrestricted
  post--{H}artree--{F}ock calculations of effective exchange integrals in Ising
  and Heisenberg models. \emph{Chem. Phys. Lett.} \textbf{1993}, \emph{210},
  201--210\relax
\mciteBstWouldAddEndPuncttrue
\mciteSetBstMidEndSepPunct{\mcitedefaultmidpunct}
{\mcitedefaultendpunct}{\mcitedefaultseppunct}\relax
\EndOfBibitem
\bibitem[Cole \latin{et~al.}(1985)Cole, Purvis, and Bartlett]{cole1985}
Cole,~S.~J.; Purvis,~G.~D.; Bartlett,~R.~J. Singlet-triplet energy gap in
  methylene using many-body methods. \emph{Chem. Phys. Lett.} \textbf{1985},
  \emph{113}, 271 -- 274\relax
\mciteBstWouldAddEndPuncttrue
\mciteSetBstMidEndSepPunct{\mcitedefaultmidpunct}
{\mcitedefaultendpunct}{\mcitedefaultseppunct}\relax
\EndOfBibitem
\bibitem[Standard(2016)]{standard2016effects}
Standard,~J.~M. Effects of Solvation and Hydrogen Bond Formation on Singlet and
  Triplet Alkyl or Aryl Carbenes. \emph{J. Phys. Chem. A} \textbf{2016},
  \emph{121}, 381--393\relax
\mciteBstWouldAddEndPuncttrue
\mciteSetBstMidEndSepPunct{\mcitedefaultmidpunct}
{\mcitedefaultendpunct}{\mcitedefaultseppunct}\relax
\EndOfBibitem
\bibitem[Finley \latin{et~al.}(1998)Finley, Malmqvist, Roos, and
  Serrano-Andr{\'e}s]{finley1998multi}
Finley,~J.; Malmqvist,~P.-{\AA}.; Roos,~B.~O.; Serrano-Andr{\'e}s,~L. The
  multi-state CASPT2 method. \emph{Chem. Phys. Lett.} \textbf{1998},
  \emph{288}, 299--306\relax
\mciteBstWouldAddEndPuncttrue
\mciteSetBstMidEndSepPunct{\mcitedefaultmidpunct}
{\mcitedefaultendpunct}{\mcitedefaultseppunct}\relax
\EndOfBibitem
\bibitem[Koopmans(1934)]{koopmans1934zuordnung}
Koopmans,~T. {\"U}ber die Zuordnung von Wellenfunktionen und Eigenwerten zu den
  einzelnen Elektronen eines Atoms. \emph{Physica} \textbf{1934}, \emph{1},
  104--113\relax
\mciteBstWouldAddEndPuncttrue
\mciteSetBstMidEndSepPunct{\mcitedefaultmidpunct}
{\mcitedefaultendpunct}{\mcitedefaultseppunct}\relax
\EndOfBibitem
\end{mcitethebibliography}

\end{document}


\section{Additional results}\label{SI_sec:results}


\begin{figure}[h]
    \centering
    \includegraphics[scale=0.5]{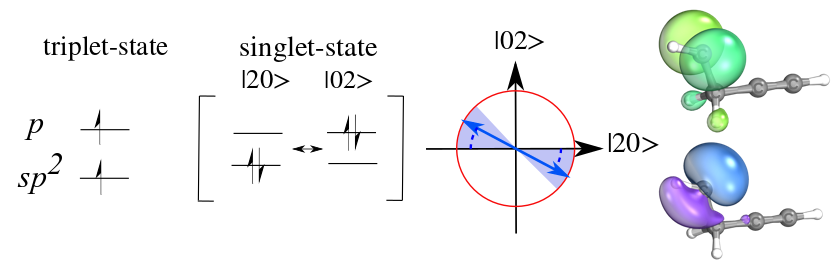}
    \caption{Left: Electronic configuration of carbenes for triplet and singlet state; middle: Schematic illustration of the singlet state orbital entanglement angle formed by the configurations state |20$\rangle$ and |02$\rangle$; right: CASSCF(2o,2e) active space orbitals of an exemplary carbene molecule.   }
    \label{fig:el_conf}
\end{figure}

\begin{figure}
    \centering
    \includegraphics[scale=0.55]{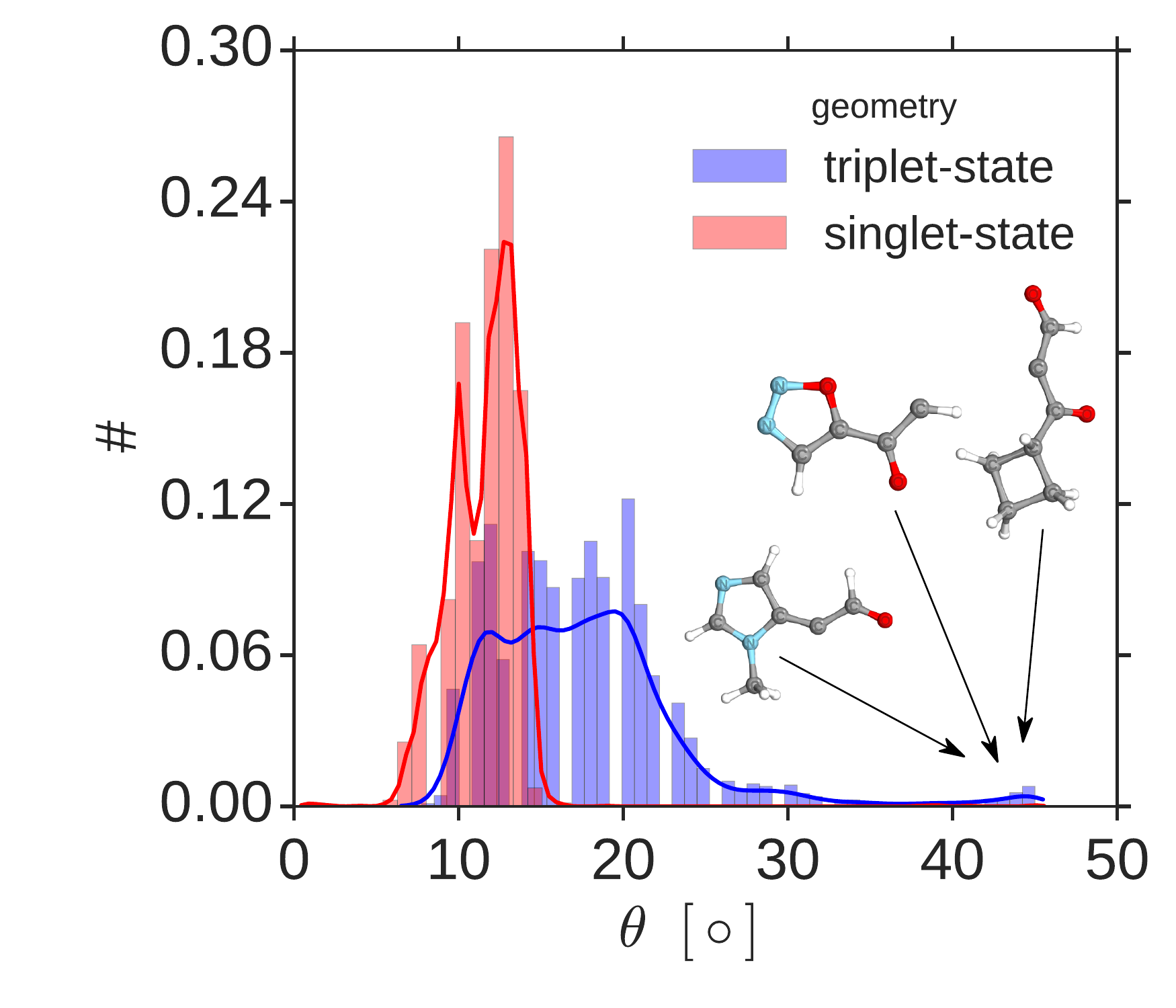}
    \caption{Distribution of the range of entanglement angle in the carbene data set for singlet-state CASSCF/\!/cc-pVDZ-F12 optimized geometries (red) and triplet-state B3LYP optimized geometries (blue). }
    \label{fig:ci_ang_distr}
\end{figure}

\newpage
\begin{table}[t]
    \centering

    \begin{tabular}{c|c|c|c|c}
    \multicolumn{5}{c}{Extreme low values of $\Delta E_{\rm{s-t}/\!/t}$}\\ \hline
    T1 & T2 & T3 & T4 & T5 \\
    \hline
        \includegraphics[scale=0.28]{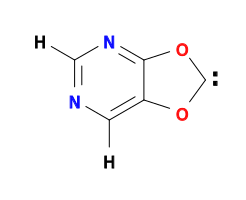} & \includegraphics[scale=0.28]{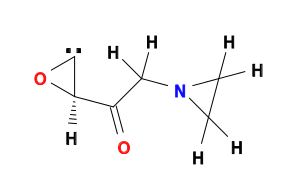} & \includegraphics[scale=0.28]{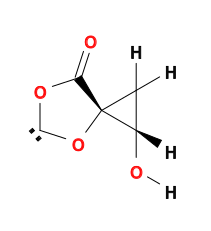} &
       \includegraphics[scale=0.28]{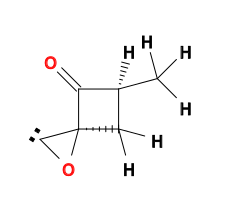}  &
        \includegraphics[scale=0.3]{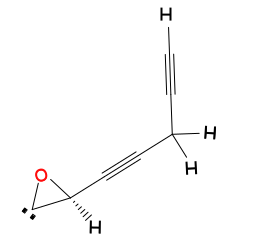} \\
        -58.3 &   -53.9 &  -53.5  &  -51.6 &   -51.6  \\
        \hline
        \multicolumn{5}{c}{\phantom{YYYY}}\\
       \multicolumn{5}{c}{Extreme high values of $\Delta E_{\rm{s-t}/\!/t}$ (monosubstituted methylidenes)}\\
        \hline
       T6 & T7 & T8 & T9 & T10 \\
    \hline
       \includegraphics[scale=0.28]{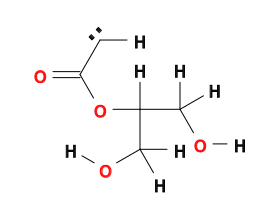} &
       \includegraphics[scale=0.28]{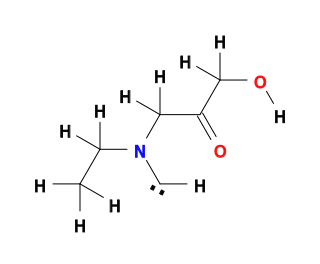} &
       \includegraphics[scale=0.3]{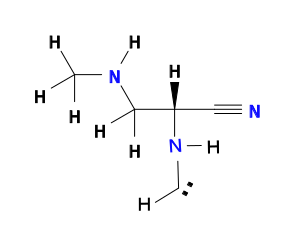} &
       \includegraphics[scale=0.3]{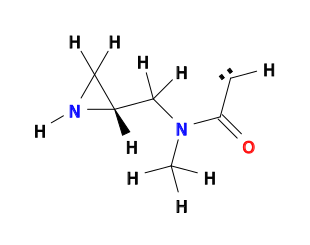} & 
       \includegraphics[scale=0.28]{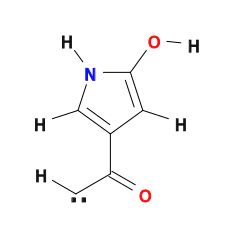}  \\
         29.8 &  29.5 &  29.4 &  29.3  &   29.2 \\ 
          \hline
         \multicolumn{5}{c}{\phantom{YYYY}}\\
        \multicolumn{5}{c}{Extreme high values of $\Delta E_{\rm{s-t}/\!/t}$ (disubstituted methylidenes)}\\
        \hline  
          T11 & T12 & T13 & T14 & T15 \\
    \hline 
      
       \includegraphics[scale=0.3]{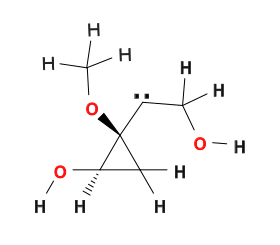} &
       \includegraphics[scale=0.3]{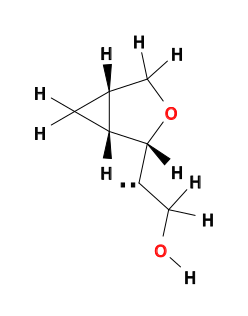} &
        \includegraphics[scale=0.3]{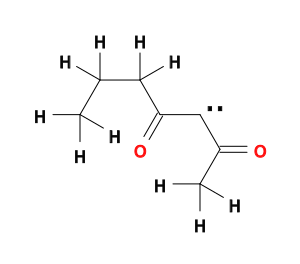} &
          \includegraphics[scale=0.3]{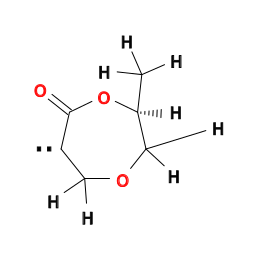} &
         \includegraphics[scale=0.3]{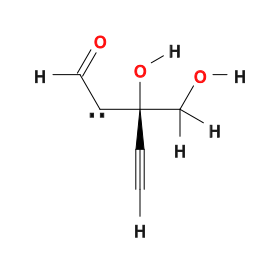} \\
          27.6  &  27.5 &
           26.2  &  25.4 &
          25.0\\ 
          \hline
          T16 & T17 & T18 & T19 & T20 \\
    \hline 
       \includegraphics[scale=0.3]{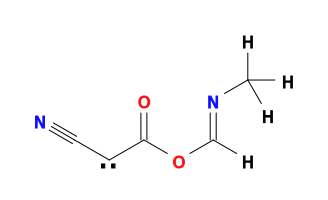}  &
        \includegraphics[scale=0.3]{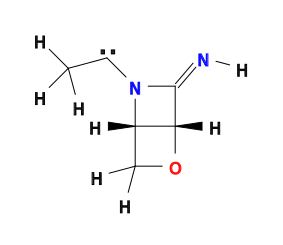} &
        \includegraphics[scale=0.3]{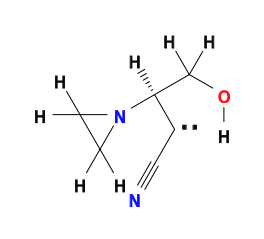} &
         \includegraphics[scale=0.3]{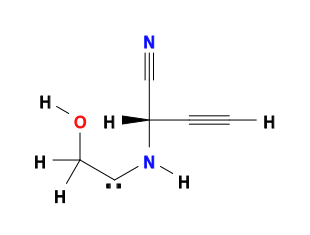} &
        \includegraphics[scale=0.3]{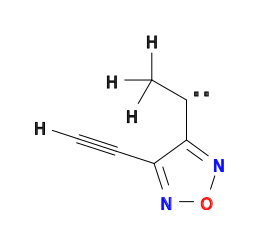} \\
         24.4  &  24.1 &   20.2 &  21.9  &  20.5  \\ 
            \hline
      
    \end{tabular}
    \caption{Extreme carbenes with lowest singlet (top) and triplet (bottom) state energy $\Delta E_{\rm{s-t}/\!/t}$ calculated at MRCISD+Q-F12/cc-pVDZ-F12//B3LYP/def2-TZVP level of theory.}
    \label{tab:type_0_t}
\end{table}

\newpage
\begin{table}[t]
    \centering

    \begin{tabular}{c|c|c|c|c}
    \multicolumn{5}{c}{Extreme low values of $\Delta E_{\rm{s-t}/\!/s}$}\\
    \hline
     S1 & S2 & S3 & S4 & S5 \\
    \hline
        \includegraphics[scale=0.28]{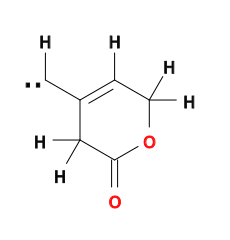} & \includegraphics[scale=0.28]{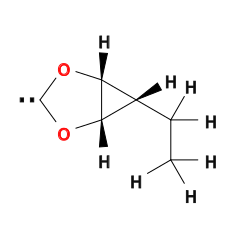} & \includegraphics[scale=0.28]{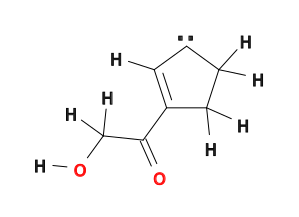} &
        \includegraphics[scale=0.3]{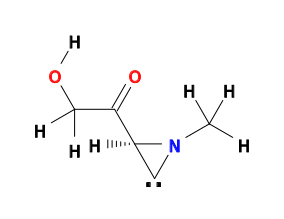} &
       \includegraphics[scale=0.28]{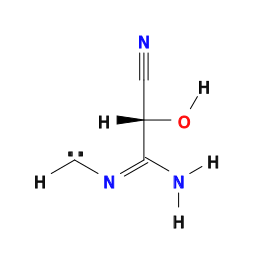}  
         \\
       $-76.7 $  & $-77.3 $   & $-77.3$   & $-87.3 $  & $-82.7  $    \\
        \hline
        
        \includegraphics[scale=0.28]{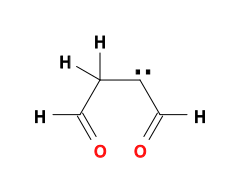} & \includegraphics[scale=0.28]{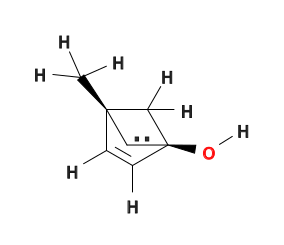} &
        \includegraphics[scale=0.28]{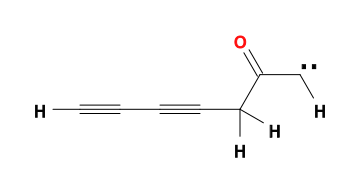} &
       \includegraphics[scale=0.28]{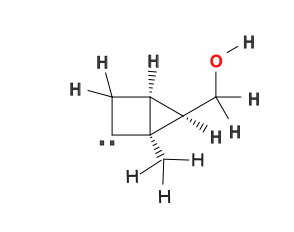}  &
        \includegraphics[scale=0.3]{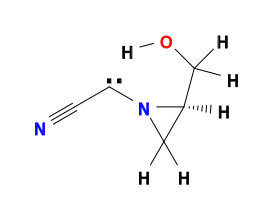} \\
       $-108.1 $  & $-113.7 $  & $-115.1 $   & $-126.0 $  & $-132.4 $   \\
        \hline

       \multicolumn{5}{c}{Extreme high values of $\Delta E_{\rm{s-t}/\!/s}$}\\
        \hline
      
       \includegraphics[scale=0.28]{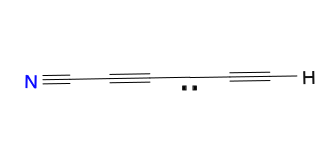} &
       \includegraphics[scale=0.28]{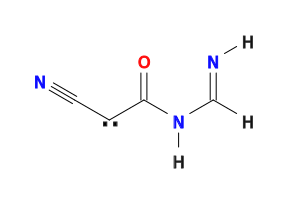} &
        \includegraphics[scale=0.28]{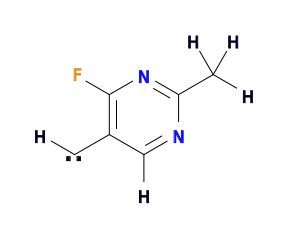} &
       \includegraphics[scale=0.3]{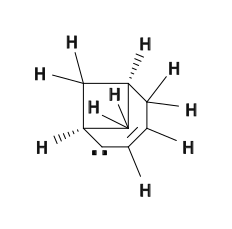} &
       \includegraphics[scale=0.3]{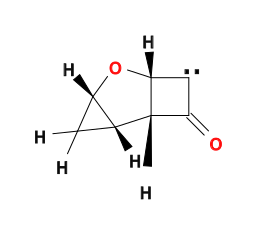} 
        \\
        $13.3 $ & $6.1 $ & $3.2$  &  $1.1 $  & $ 1.1 $    \\

    \end{tabular}
    \caption{Extreme carbenes with lowest singlet (top) and triplet (bottom) state energy $\Delta E_{\rm{s-t}/\!/s}$ calculated with MRCISD+Q/cc-pVDZ-F12//CASSCF/cc-pVDZ-F12}
    \label{tab:type_0_s}
\end{table}



\begin{figure}
    \centering
    \includegraphics[scale=0.4]{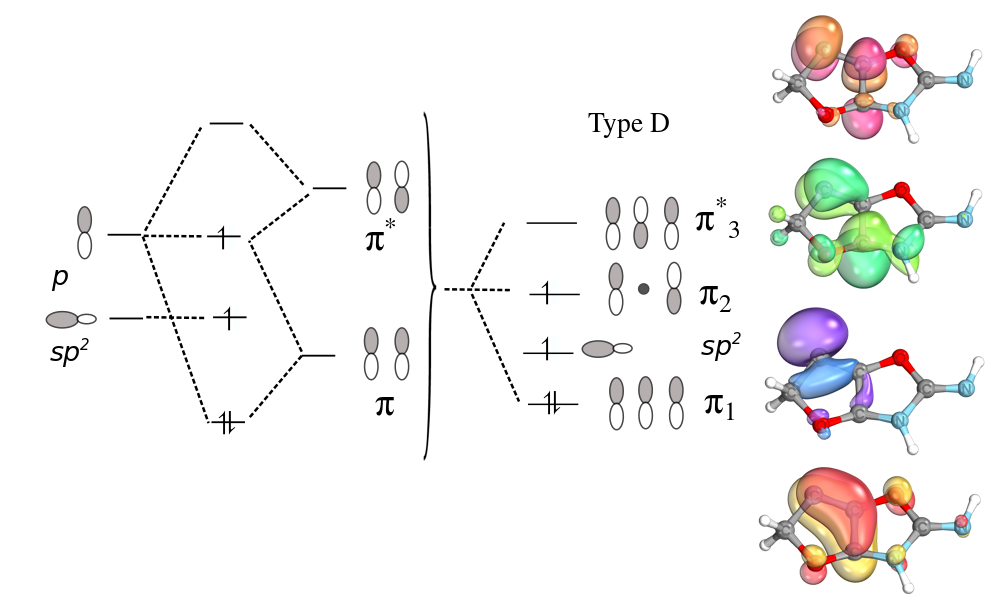}
    \caption{A molecular orbital diagram of a carbene center coupled to an electrophilic $\pi$-system along with an example molecule that has been sorted out from the data set due to a strongly delocalized carbene nonbonding orbital.}
    \label{fig:gdb9_024972_0}
\end{figure}

   \begin{table}[t]
    \centering
   \begin{tabular}{c|c|c|c|c}
        \hline
       \includegraphics[scale=0.28]{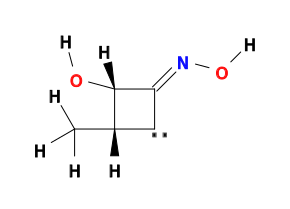} &
       \includegraphics[scale=0.28]{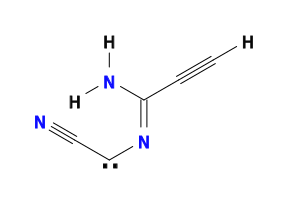} &
       \includegraphics[scale=0.28]{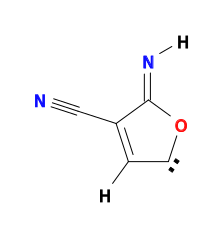} &
       \includegraphics[scale=0.28]{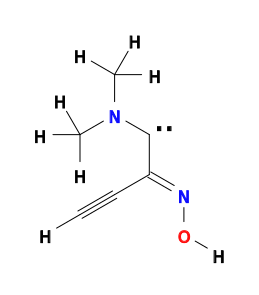} &
       \includegraphics[scale=0.28]{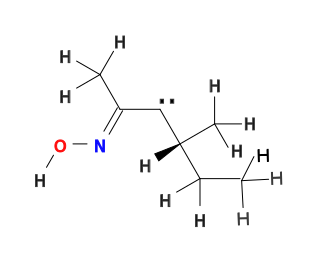} 
       \\
        S0 $\quad$ 22.5  & S0 $\quad$ 15.6  & S0 $\quad$ 7.6 & S0 $\quad$ 17.6 & S0 $\quad$ 18.4 \\ 
        S1 $\quad$ 31.3 & S1 $\quad$ 31.1 & S1 $\quad$ 19.7 & S1 $\quad$ 41.1 & S1 $\quad$ 35.6 \\ 
        \hline
    \includegraphics[scale=0.28]{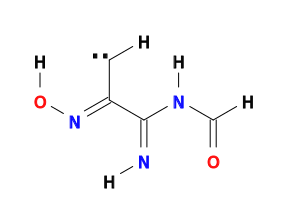} &
     \includegraphics[scale=0.28]{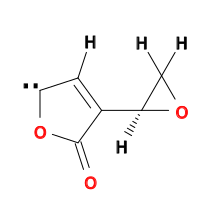} &
     \includegraphics[scale=0.28]{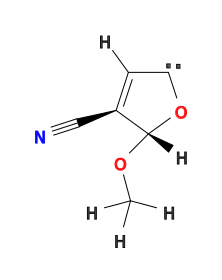} &
      \includegraphics[scale=0.28]{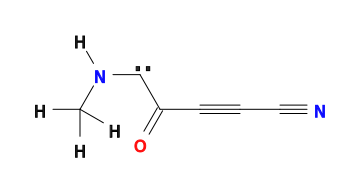} &
      \includegraphics[scale=0.28]{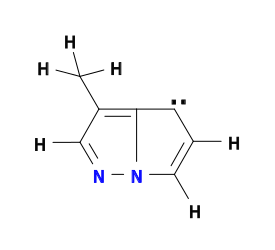} 
        \\
        S0 $\quad$ 24.5 & S0 $\quad$ 8.7 & S0 $\quad$ 2.9 & S0 $\quad$ -3.0 & S0 $\quad$ 19.9 \\ 
        S1 $\quad$ 39.1 & S1 $\quad$ 20.6 & S1 $\quad$ 19.9 & S1 $\quad$ 22.8  & S1 $\quad$ 21.5 \\ 
        \hline
    \end{tabular}
    \caption{Example set of carbene centers coupled to a strongly electrophilic $\pi$-system. Spin gap energies $\Delta E_{\rm{s-t}/\!/t}$ given with respect to the lowest triplet state ``$t$'' and the two lowest carbene singlet states ``S0'' and ``S1'' are given in kcal/mol.}
    \label{tab:type_D}
\end{table}
      

\begin{table}[]
    \centering
    \begin{tabular}{c|c|c|c}
     Types of triplet configuration  &  \includegraphics[scale=0.28]{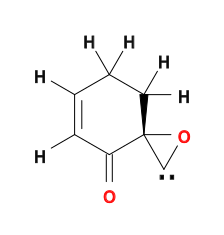}  
       &  \includegraphics[scale=0.28]{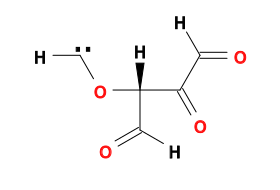}  
       &  \includegraphics[scale=0.28]{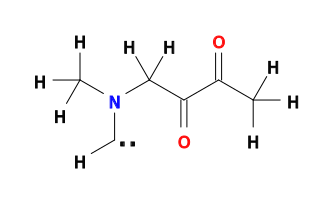} 
         \\
         \hline
       \includegraphics[scale=0.4]{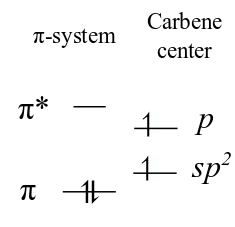}  & 76\% & 96\% & 97\%  \\ \hline
       \includegraphics[scale=0.37]{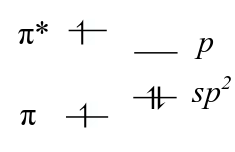} & 82\%$^{a}$ & - & -  \\ \hline
       \includegraphics[scale=0.37]{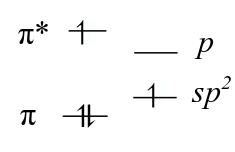} & 13\%$^{a}$ & 4\% & 3\% \\ \hline
       \includegraphics[scale=0.37]{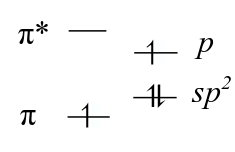}  & 16\% & - & - \\ \hline
         & & & \\ 
     $\Delta E_{\rm{s-t}/\!/t}$   & -52.3 & 11.0 & 26.8\\
       &  & & \\    & -96.1$^{a}$ & - & -\\ \hline
    \end{tabular}
    \caption{Most important triplet state configurations and their weight in the CASSCF wave function for exemplary carbenes with orbitals on a neighbouring functional group interacting with the carbene center non-bonding orbitals via strong hyperconjugation. The contribution of each type of configuration within an SA-CASSCF(4e,4o) calculation including the four lowest triplet and the two lowest singlet states is given in percentage. The singlet-triplet gap energies are given in kcal/mol for the lowest states as calculated with MRCISD+Q-F12/cc-pVDZ-F12. Superscript $^a$ indicates the first excited triplet state.  }
    \label{tab:type_1_2}
\end{table}


\clearpage

\section{The electronic structure of carbenes} \label{SI_sec:multi-ref-theory}

Strong electron correlation can play an important role in the carbene singlet state as the non-bonding orbitals on the carbene center may be close in energy and are close in space by definition.
Hence, the essential physical interactions relevant for the spin gap may be described by treating the two non-bonding orbitals at the carbene center within a complete active space (CAS) calculation.
In fact, it is known since the early 1980's \cite{feller1982theoretical} that only the inclusion of both non-bonding orbital closed shell configuration ``$|20\rangle$'' and ``$|02\rangle$'' in the SCF method yields qualitatively correct singlet-triplet gaps ($\approx$10 kcal/mol) for methylene with respect to experiment (9.08 kcal/mol $\pm$ 0.18). The basis set converged HF spin gap is near 25 kcal/mol).\cite{shavitt1985geometry}
We advocate that this two-configuration character of the singlet state wave function (WF) should reflect in the carbene MO diagram\cite{irikura1992singlet} whenever strong correlation is significant.  
Figure \ref{fig:el_conf} depicts the corresponding carbene MO diagram.

A balanced and qualitatively correct treatment of both spin states may then in the general case be obtained by a state-averaged (SA) CASSCF(2e,2o) calculation.
For the triplet state of the carbene, the SA-CASSCF wave function has a single-reference character 
$\big(\Psi_{S=1}^{\rm{CASSCF}} = \mathcal{A}|\phi_i^{\alpha}, \phi_i^{\beta}, \phi_j^{\alpha}, \phi_j^{\beta},$\ldots $,\phi_r^{\alpha}, \phi_s^{\alpha}|$, $E_{s=1} = \langle{\Psi_{S=1}^{\rm{CASSCF}}}| \hat{H}| {\Psi_{S=1}^{\rm{CASSCF}}}\rangle\big)$, 
where $\phi_i,\phi_j,\dots$ are spin-restricted closed-shell orbitals with the corresponding spin function, $r,s$ are spin-restricted active orbitals, mainly located on the carbene center, and $\mathcal{A}|\ldots|$ denotes a Slater-Determinant.

For the singlet state ($S=0$), the two closed-shell configurations $R$ and $S$ are the ones obtained for natural orbitals:
\begin{equation}
\begin{split}
    \Psi_{S=0}^{\rm{CASSCF}} = c_{R}\mathcal{A}|\phi_i^{\alpha}, \phi_i^{\beta}, \phi_j^{\alpha}, \phi_j^{\beta},\ldots, \phi_r^{\alpha},\phi_r^{\beta}|\\
    +  c_{S}\mathcal{A}|\phi_i^{\alpha}, \phi_i^{\beta}, \phi_j^{\alpha}, \phi_j^{\beta},\ldots, \phi_s^{\alpha},\phi_s^{\beta}|,
\end{split}
\label{eq:singwavefunc}
\end{equation}
where $c_T$ and $c_U$ are the complex-valued configuration interaction (CI) coefficients. 
The singlet state energy then reads:
\begin{equation}
\begin{split}
        E^{\rm{CASSCF}}_{S=0} = |c_R|^2E_{S=0,R} + |c_S|^2E_{S=0,S} \\
       + (c^*_Rc_S + c^*_Sc_R)K_{rs} \quad .
\end{split}
\label{eq:casscf_en}
\end{equation}
$E_{S=0,R}$ and $E_{S=0,S}$ are the mean field energy contributions of the configurations $R$ and $S$, respectively.
$K_{rs} = (rs|rs)$ (Mulliken notation) is an exchange interaction that arises from the strong correlation of the two electrons in the active orbitals and that stabilizes the singlet state. 
The strength of the strong correlation can be expressed as an angle $\theta = \arctan\left(\frac{|c_S|}{|c_R|}\right), 0^{\circ} \le \theta \le 45^{\circ}$.
For $\theta$ larger than a few degrees, the system is generally considered of significant multireference character. 

Early computations at the MRCISD+Q level of theory with a two-configuration SCF wave function were able to reproduce the methylene singlet-triplet spin gap up to a few tenths of kcal/mol when approaching the basis set limit\cite{shavitt1985geometry,bauschlicher19871}. 

The MRCISD method introduces dynamic correlation on top of the CASSCF reference wave function by including singly and doubly excited configurations in the variational optimization of the wave function.
Empirical quadruple excitations contributions (``+Q'') are included via the Davidson correction.\cite{langhoff1974configuration}
Including singles and doubles excitations in the quasi-variational way of internally contracted MRCI allows for a certain ``flexibility'' in the wave function to properly describe the diverse carbene systems in our chemical space.

Methods based on the configuration interaction expansion may suffer from important size-consistency errors, even though the Davidson correction generally alleviates this problem.
This may for example lead to noticeable errors in the balanced description of effective spin couplings in carbene systems along dissociation curves.\cite{yamaguchi1993comparison} 
However, singlet-triplet gaps in carbene monomers can be viewed as a rather size-intensive property and therefore the size-consistency defect of MRCISD+Q is much less significant.
Even though the coupled cluster (CC) methods are in general the most reliable way to compute accurate dynamical correlation contributions, computationally affordable single-reference CC methods do not perform systematically better than MRCISD+Q in the case of carbenes. \cite{cole1985,standard2016effects}
Single-reference post-HF methods depend strongly on the quality of HF as a zeroth-order approximation, we therefore opted to use MRCISD+Q as our benchmark reference.

\section{Qualitative assessment of $\mathbf{\Delta E}_{\rm{s-t}}$}
\label{SI_sec:derivation}

Using the singlet and triplet state wave functions and energy expressions of SA-CASSCF(2e,2o), as described in Section \ref{SI_sec:multi-ref-theory} in the SI, the vertical spin gap $\Delta E_{\rm{s-t}}^{\rm{CASSCF}}$ for a given carbene structure can be written as (assuming real-valued configuration interaction coefficients):

\begin{equation}
\begin{split}
    \Delta E_{\rm{s-t}}^{\rm{CASSCF}} &= \Big(1+ \underbrace{\hat{P}_{rs}\hat{P}_{RS}}_{\text{operators that permute $r/R$ and $s/S$}}\Big) \Big[ \big[ 2|c_R|^2 - 1 \big] \big[\underbrace{k_{rr} + v_{rr}}_{\text{1-electron interactions}} +\\
    &\underbrace{\sum_i (2J_{ir} - K_{ir})}_{\text{mean field interaction with the closed shell electrons}} \big] + \underbrace{|c_R|^2 J_{rr}}_{\text{active orbital Coulomb interaction $S=0$}} \Big] - \\
    & \underbrace{\big( J_{rs} - K_{rs} \big)}_{\text{active orbital Coulomb and exchange interaction $S=1$}} \qquad -
    \underbrace{\big[ 2|c_R c_S| K_{rs}\big]}_{\text{active orbital resonance energy $S=0$}}
\end{split}
\label{eq:e_st_equation}
\end{equation}
where the index $i$ runs over closed shell orbitals and indices $r$ and $s$ over active orbitals of the carbene and $k_{rr}$ and $v_{rr}$ are the kinetic and electron-nuclei attraction energies, respectively.
The Coulomb interaction $J_{rs}$ term reads $(rr|ss)$ in Mulliken notation.
For convenience, all effective 1-electron interactions of an active orbital can be grouped into one variable $\varepsilon^{(1)}_r = k_{rr} + v_{rr} + \sum_i{(2J_{ir} - K_{ir})}$.
Introducing the variable $\alpha = 2|c_R|^2 - 1$ and using $|c_R|^2 + |c_S|^2 = 1$, eq. \ref{eq:e_st_equation} can be rewritten as:

\begin{equation}
    \Delta E_{\rm{s-t}}^{\rm{CASSCF}}  = \alpha \Big(1 -\hat{P}_{rs}\Big)\Big(\varepsilon^{(1)}_r + \frac{1}{2}J_{rr} \Big) + J_{rs}^+ - J_{rs} + \big( 1- \sqrt{1-\alpha^2} \big)K_{rs}
    \label{eq:split1}
\end{equation}
where the active orbital Coulomb repulsion terms have been regrouped according to their dependence on $\alpha$ with $J_{rs}^+ = \frac{1}{2}\big(  J_{rr} +  J_{ss} \big)$.

Another instructive way of writing the SA-CASSCF(2e,2o) energy expression for the singlet or triplet state is in terms of general one-particle and two-particle reduced density matrices in the active space spinorbital basis:
\begin{equation}
    E^{\rm{CASSCF}}=E_c+ E_{\rm{nuc}}+\sum_{rs} f^c_{rs} D^{(1)}_{rs}+ \sum_{rstu} D^{(1)}_{rs} \left[ (rs|tu) - (ru|ts) \right] 
    + \mathcal{O}(D^{(2)}_{rs,tu})
    \label{eq:densmat}
\end{equation}
 where $f^c_{rs}=\delta_{rs}\varepsilon^{(1)}_r$ is the closed shell orbital Fock operator and $E_c$ and $E_{\rm{nuc}}$ are the closed shell energy and nuclear interaction, respectively.
 The only contribution that is essentially expressed in terms of the active space two-particle reduced density matrix $D^{(2)}_{rs,tu}$ is the singlet state resonance energy from the strong correlation.
 All other terms in eq. \ref{eq:densmat} can be expressed as the CASSCF wave function expectation values of a generalized Fock operator that takes into account that the mean field density of active space orbital Coulomb and exchange interactions is expressed by more than one reference configuration.\cite{finley1998multi}
 As we are using natural orbitals (NO) in the active space, the state-averaged one-particle reduced density matrix take simple diagonal forms and the generalized Fock operator representation in the NO active orbital basis $\mathbf{F}^{\rm{NO}}$ is in general diagonal dominant.
 Given that we verify that the natural orbitals active space reproduces well the concept of nonbonding carbene orbitals, the natural orbital energies may be an intuitive way of quantifying the concept of energy splitting of the carbene nonbonding orbitals.
 Since the active space orbitals $r$ and $s$ are at least partially occupied in the CASSCF generalized mean field, from the perspective of Koopmans' theorem,\cite{koopmans1934zuordnung} these orbital energies may be seen as state-averaged formal negative ionization energies of the orbital associated singlet and triplet state configurations $c_R$ and $c_S$, i. e.
 \begin{equation}
  \varepsilon_r = \varepsilon^{(1)}_r + \frac12 J_{rs} -\frac12 K_{rs} + \frac12 J_{rr}\ , \quad \varepsilon_s=\hat{P}_{rs} \varepsilon_r  \quad .
 \end{equation}
 It should be noted that the diagonal elements of $\mathbf{F}^{\rm{NO}}$ would be a less suited choice for the definition of orbital energies given that the singlet state contribution to the state averaged generalized Fock operator will converge to the single-reference Fock operator for $\theta\to0$. In that case the orbital $s$ would be treated as a virtual orbital and the orbital energy would not be a meaningful approximation of the negative ionization energy.
 The contributions linear in $\alpha$ in eq. \ref{eq:split1} can then be approximated by the carbene orbital energy splitting $\Delta \varepsilon_{rs} = \varepsilon_r  - \varepsilon_s $.
 
Furthermore, from the definition of $\alpha$ it follows that the entanglement angle $\theta$ can be written as $\theta = \frac{1}{2}\arccos(\alpha)$.
Finally, expressing the spin gap in terms of $\theta$ then yields:
\begin{equation}
\begin{split}
    \Delta E_{\rm{s-t}}^{\rm{CASSCF}} = cos(2\theta)\ \Delta \varepsilon_{rs} + \big[ 1- sin(2\theta)\big]K_{rs} + \Delta J_{rs}^+ 
\end{split}
\label{eq:e_st_eq_short}
\end{equation}
with $\Delta J_{rs}^+ = J_{rs}^+ - J_{rs}$.
$J_{rs}^+$ can be understood as the orbital-averaged Coulomb interaction in the active space for the singlet state. Since the electrons occupy the same active orbital in the singlet state configurations, $J_{rs}^+$ is in general significantly larger in magnitude than the corresponding triplet state Coulomb repulsion $J_{rs}$ and hence $\Delta J_{rs}^+>0$. 



In the regime of strong orbital entanglement ($\theta \ge 25^{\circ}$) a second order Taylor expansion of eq. \ref{eq:e_st_eq_short} at $\theta=45^{\circ}$ yields: 
\begin{equation}
\begin{split}
     \Delta E_{\rm{s-t}}^{\rm{CASSCF}} \approx 2\Delta\theta \Delta \varepsilon_{rs} +  2\Delta\theta^2 K_{rs} + \Delta J_{rs}^+ 
\end{split}
\label{eq:e_st_eq_short2}
\end{equation}
where $\Delta\theta = 45^{\circ} - \theta$.
In the regime of moderate orbital entanglement ($\theta \le 15^{\circ}$) a second order Taylor expansion of eq. \ref{eq:e_st_eq_short} at $\theta=0^{\circ}$ yields:
\begin{equation}
\begin{split}
    \Delta E_{\rm{s-t}}^{\rm{CASSCF}} \approx \big[1-2\theta^2\big]\ \Delta \varepsilon_{rs} + \big[ 1- 2\theta\big]K_{rs} + \Delta J_{rs}^+ 
\end{split}
\label{eq:e_st_eq_short3}
\end{equation}
Partial differentiation of $\Delta E_{\rm{s-t}}^{\rm{CASSCF}}$ with respect to $\theta$ and setting $\frac{\partial \Delta E_{\rm{s-t}}(\theta)}{\partial \theta}=0$ yields expressions for $\theta_{\mathrm{min}}$ which can be inserted into eqs.\ \ref{eq:e_st_eq_short2} and \ref{eq:e_st_eq_short3}, yielding the equations given in the main text.

\begin{figure}
    \centering
    \includegraphics[scale=0.4]{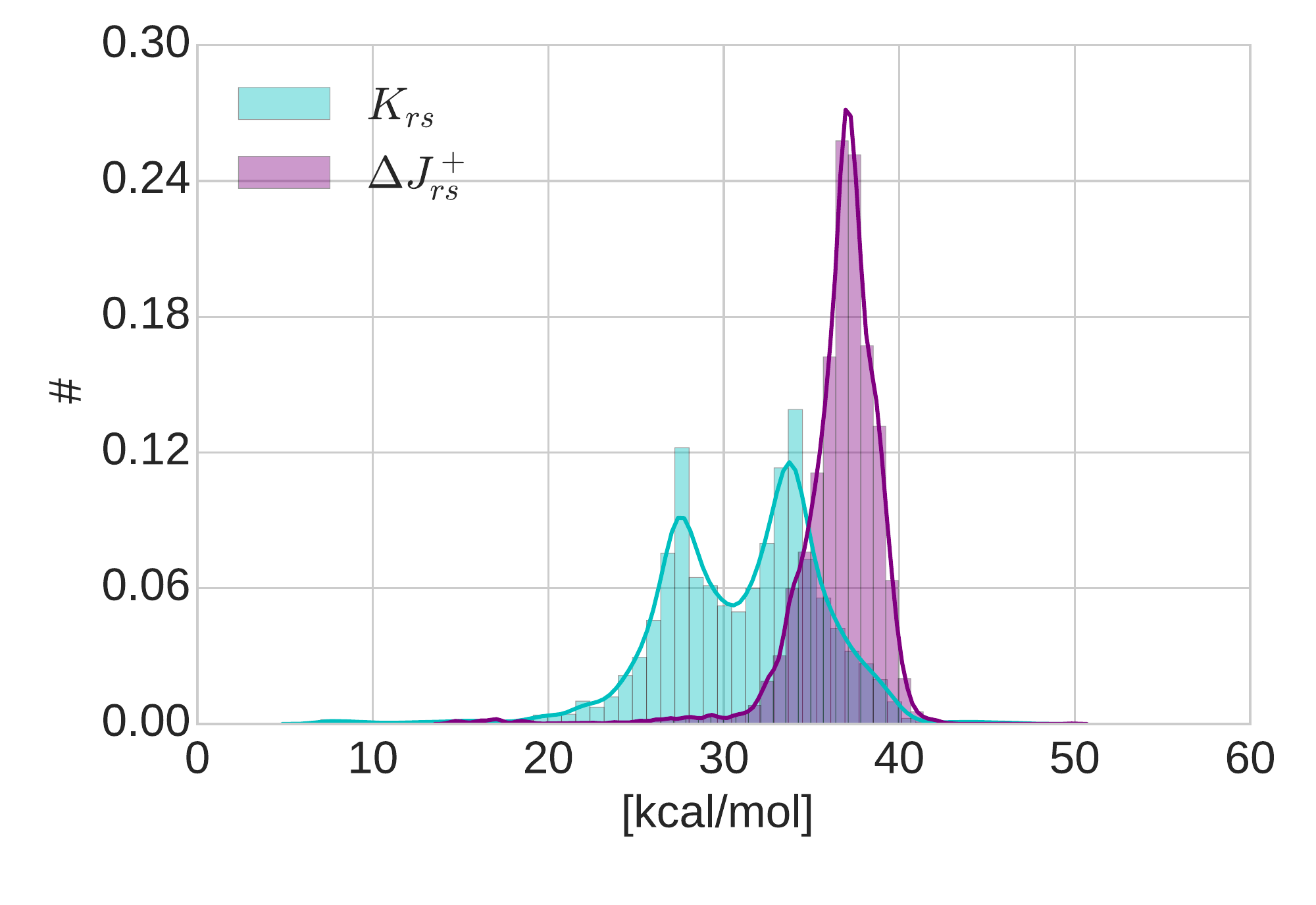}
    \caption{Distribution of the Coulomb $\Delta J_{rs}^+$ and the exchange $K_{rs}$ integrals calculated using SA-CASSCF(2e,2o)/pVDZ-F12. }
    \label{fig:Jrs_Krs}
\end{figure}


\bibliographystyle{ieeetr}
\bibliography{references}